\documentclass{openjournal}

\usepackage{xcolor}
\usepackage{textgreek}
\usepackage[utf8]{inputenc}
\usepackage[english]{babel}

\usepackage{hyperref}
\hypersetup{
    unicode, 
    colorlinks=true,
    linkcolor=linkcolor,
    citecolor=linkcolor,
    filecolor=linkcolor,
    urlcolor=linkcolor,
}
\usepackage{color,colortbl}
\definecolor{linkcolor}{rgb}{0.0,0.3,0.5}
\usepackage{tensind}
\tensordelimiter{?}
\DeclareGraphicsExtensions{.bmp,.png,.jpg,.pdf}
\usepackage{verbatim}
\usepackage[normalem]{ulem}
\usepackage{orcidlink}

 \usepackage{graphicx}
 \usepackage{amsmath}
 \usepackage{amssymb}
 \usepackage{amsthm}
 \usepackage[utf8]{inputenc}
 \usepackage{slashed}

 \usepackage{mathrsfs}
 \usepackage{physics}
 \usepackage[compat=1.0.0]{tikz-feynman}
 \usepackage{tikz}
 
\usepackage{tikz-feynman}

\expandafter\let\csname ver@longtable.sty\endcsname\relax
\usepackage{subfig}

\graphicspath{ {./figs/} }

\begin{document}

\title{Cosmological Dynamics of Multi-Axion Quintessence}

\author{Supakorn Katewongveerachart}
\affiliation{Department of Physics, Faculty of Science, Mahidol University, Bangkok, Thailand}
\author{David J.E. Marsh}
\affiliation{Department of Physics, Faculty of Natural, Mathematical and Engineering Sciences, King’s College London, UK}

\begin{abstract}
   One fundamental physics interpretation of Dark Energy Spectroscopic Instrument (DESI) results is that the observed accelerated expansion of the Universe is driven not by the cosmological constant, but by a slowly rolling scalar field, a natural model for which is an axion. 
   In the ``string axiverse'' one expects not one, but many, axions. 
   We thus investigate dark energy dynamics in a models with two, three, and four axions, and compute the prior probability for the dark energy equation of state parameters $(w_0,w_a)$ implied by priors on the fundamental parameters of the model conditioned weakly on the resulting cosmology. We find various interesting behaviours, for example where multiple fields can be in an oscillating regime while maintaining $w_0<-1/3$, and models with opposite sign for $w_a$ than in a single-field model. The prior probability in $(w_0,w_a)$ gains support away from the ``thawing quintessence'' behaviour preferred by DESI. Similarly, adding more axion fields and interactions among them leads to the prior support in the baryon acoustic oscillation distance measures moving further from the DESI data. Thus, interacting multifield models will not generically be preferred over simpler models unless very exotic dark energy evolution happens to be observed in future. The novel equations of state we illustrate may, however, be of relevance to models of early dark energy or inflation/reheating with multiple axions, and further study of multifield dynamics is warranted.

\end{abstract}


\maketitle

\section{Introduction}

The concordance cosmological model has, since at least 1998~\citep{SupernovaSearchTeam:1998fmf,SupernovaCosmologyProject:1998vns}, pointed to a Universe undergoing accelerated expansion at present, i.e. Friedmann-Robertson-Walker (FRW) scale factor $\ddot{a}>0$. The simplest theoretical explanation for this fact is that the cosmological constant $\Lambda>0$. Alternative ``dark energy'' (DE) models can be parameterised by an equation of state, $w = P/\rho$, with $P$ and $\rho$ the DE fluid pressure and energy densities, with $w_\Lambda=-1$ a constant. Until recently there has been no preference for any deviation from the predictions of the $\Lambda$ model, e.g. in the cosmic microwave background (CMB) anisotropies~\citep{Planck:2018vyg}.

The Dark Energy Spectroscopic Instrument (DESI) measures the baryon acoustic oscillations (BAO) in galaxy clustering, which, in combination with the CMB, provide a local distance measurement that can be used to precisely test models of DE ~\citep{DESI:2024mwx}. Surprisingly, this analysis now favours, with significance in the 2-3$\sigma$ range, a departure from $w_\Lambda=-1$, signaling that the DE density may be evolving in time \citep[although see more recent reassessments such as][that show a reduced significance]{Ong:2026tta}. A simple parameterisation of this evolution is the \cite{Chevallier:2000qy,Linder:2002et} (CPL) Taylor expansion: $w(a)=w_0 +(1-a)w_a$, where $a=1$ is normalised to the present day. In terms of this parameterisation, DESI data are best fit by $w_0>-1$ and $w_a<0$, with the precise values depending on the combination of datasets. 

A simple model for time evolving DE is a single homogeneous and slowly rolling scalar field, $\phi$, with potential energy $V(\phi)$, and a canonical kinetic energy \citep[reviews of alternatives can be found in][]{Copeland:2006wr,Clifton:2011jh,Bull:2015stt}. Accelerated expansion can be driven by the scalar field if the mass, $m_{\phi,0} \equiv \partial^2V/\partial\phi^2|_0\lesssim H_0\approx 10^{-33}\text{ eV}$. A technically natural model for such an ultralight field is a pseudo-Nambu-Goldstone boson, the simplest case of which is an axion(-like) particle, as reviewed in \cite{Marsh:2015xka}. 

The simplest single axion model has potential energy~\footnote{Note that the cosmological constant here has been fixed to precisely $\Lambda_a=m_a^2f_a^2$ to give zero potential energy in the axion vacuum, with all other contributions ignored. Scalar field DE models do not solve the cosmological constant problem~\citep{Weinberg:1988cp}.} $V(\phi)=m_a^2f_a^2 [1- \cos (\phi/f_a)]$. For such a single axion model, the theory informed prior on $(w_0,w_a)$ space aligns with ``thawing quintessence''~\citep{Marsh:2014xoa} and is thus in the region preferred by the DESI measurements. A dedicated analysis of this model~\citep{Urena-Lopez:2025rad} yields best fit parameters $\log_{10}(m_a/\text{eV})=-32.6$ and $\log_{10}(f_a/M_{pl})=-0.22$ (where $M_{pl}=1/\sqrt{8\pi G_N}$ is the reduced Planck mass) when applied to the DESI+Pantheon~\citep{Brout:2022vxf}+Union3~\citep{Rubin:2023jdq}+DESY5~\citep{DES:2024jxu} datasets. It is worth noting, however, that the best fit value of $H_0$ is lower in this data combination and model than in a model with $w_\Lambda=-1$: evolving DE seems to worsen the ``$H_0$ tension'' with local distance measurements~\citep[reviewed in][]{Knox:2019rjx}.

String theory, when compactified on Calabi-Yau manifolds down to four dimensions, gives rise to many axion fields from dimensional reduction of $p$-form gauge fields~\citep{Arvanitaki:2009fg}, and so it is natural to ask whether such an axion can explain the best fit values derived from DESI. Recently \cite{Sheridan:2024vtt} computed $(m_a,f_a)$ for a large number of Calabi-Yaus in type IIB string theory in a region of the landscape expected to give the largest values of $f_a$ at small $m_a$, and found no models with large enough $f_a$ such that an axion with $m_a\approx H_0$ can contribute significantly to the DE density today. In general, for a single axion, such a combination of $(m_a,f_a)$ appears to be in conflict with some version of the so-called ``weak gravity conjecture'' \citep[see e.g.][]{Cicoli:2021gss}. 

On the other hand, as is well known in the context of theories of inflation~\citep{Kim:2004rp,Dimopoulos:2005ac}, invoking multiple axions with nearly degenerate masses it is possible to increase the effective decay constant (i.e. the traversable distance in a unit cell in field space). Thus, in the present work, we extend the analysis of \cite{Marsh:2014xoa} to a model with two axion fields and investigate the theory informed prior on $(w_0,w_a)$ and the BAO distance measures conditioned on the axions driving accelerated expansion with broadly acceptable values of $H_0$ and $\Omega_\phi=\rho_\phi/(3H_0^2 M_{pl}^2)$. The theory parameters we sample are those of the string-inspired effective field theory, which in a model with $N=2,3,4$ axions has $N$ mass parameters and $N(N+1)/2$ effective decay constants accounting for off-diagonal instanton charges after bringing the kinetic term to canonical form. We analyse the behaviour of certain novel models that depart from the single field description, and find various exotic behaviours in the mutli field dynamics. In general we do not find it so simple to lower the overall scale of decay constants, and we show that that multi-field dynamics takes axion quintessence further from the prefereed region of $(w_0,w_a)$ space, and from teh BAO data. This is expected, since the DESI data alone are well fit by a cosmological constant with only mild preference for the slow-roll thawing $(w_0,w_a)$ models.

This paper is organized as follows. In Section \ref{Theory} we introduce the basic theoretical framework and numerical method employed. In Section \ref{results} we present and discuss the results of our numerical study. We conclude in Section~\ref{sec:conclusions}. The Appendix gives an approximate analytical model to explain some of the features we observed.

\section{Multi-Field Axion Quintessence}\label{Theory}

\subsection{The Axion Potential}

The Friedmann equation is
\begin{align}
    H^2 = \frac{1}{3M_{\mathrm{Pl}}^2}\left(\frac{\rho_{r,0}}{a^4} + \frac{\rho_{m,0}}{a^3} + \rho_{\phi}\right),
    \label{eqn:full_friedmann}
\end{align}
where $a$ is the FRW scale factor, $H=\dot{a}/a$ is the Hubble parameter, and $\rho_{r,0},\rho_{m,0}$ are the radiation and matter densities when $a=1$. The density, $\rho_\phi$, and pressure, $P_\phi$, accounting for all axion fields are given by
\begin{equation}
    \rho_\phi = \sum_i \frac{\dot{\phi_i}^2}{2}+V(\vec{\phi})\, ,\quad P_\phi = \sum_i \frac{\dot{\phi_i}^2}{2}-V(\vec{\phi})\, ,
\end{equation}
which together define the equation of state $w_\phi=P_\phi/\rho_\phi$. We have used that the kinetic term for axions is $\mathcal{L}_K = K_{ij}\partial_\mu \phi_i \partial^\mu \phi_j$. With $K_{ij}$ independent of $\phi$, there is always a change of variables that brings $K_{ij}$ to the identity, and we work in this basis (we comment on this again briefly later). The scalar potential $V(\phi)$ includes the cosmological constant, and in the context of quintessence models, we simply assume that this is zero in the vacuum, i.e. $V(\langle \vec{\phi}\rangle)=0$.

First consider a single scalar field $\phi$ minimally coupled to gravity (canonical kinetic term) with the potential
\begin{equation}
    V(\phi) = m_a^2 f_a^2 \left[1 - \cos\left(\frac{\phi}{f_a}\right)\right], \label{eqn:single_cosine}
\end{equation}
where $m_a$ is the axion mass (second derivative of the potential at the minimum) and $f_a$ is the decay constant. The Klein-Gordon equation in the FRW background is:
\begin{align}
    \ddot{\phi} + 3H\dot{\phi} + \frac{dV}{d\phi} = 0.
\end{align}

Taking the background evolution dominated by matter or radiation, i.e. $(a\propto t^p)$, there is an analytic solution for $\phi$:
\begin{align}\label{single eom sol}
    \phi(t)=a^{-3/2}(t/t_i)^{1/2}\left[C_1J_n(m_at)+C_2Y_n(m_at)\right]
\end{align}
where $n=(3p-1)/2$, $J_n$, $Y_n$ are Bessel functions of first and second kind, and $t_i$ is the initial time. The exact solution only applies so long as $\rho_\phi$ does not contribute significantly to $H$, but is nonetheless useful to understand features of the dynamics. In order for $\phi$ to itself drive the expansion of the Universe with $\rho_\phi>\rho_m>\rho_r$, requires $m_a\lesssim 1/t$ such that $w_\phi<0$ and $\rho_\phi$ can overtake the matter density by scaling as $a^{-p}$ with $p<3$. At such a time that $\rho_\phi>\rho_m$, the exact solution ceases to apply. When $m_a\gtrsim 1/t$, the field begins to oscillate. For a single field with the potential Eq.~\eqref{eqn:single_cosine}, which is harmonic at the minimum, this implies $\langle w_\phi\rangle=0$~\cite{Turner:1983he}, where $\langle\cdots\rangle$ is here the oscillation period average. For a single field, the onset of oscillations precludes the field from acting as DE with $w<-1/3$.

The simplest generalisation is to consider $N$ axions with independent cosine potentials:
\begin{equation}
    V(\phi_i) = \sum_{i=1}^N \Lambda_i^4 \left(1 - \cos\left(\frac{\phi_i}{f_i}\right)\right).
\end{equation}
We call this the non-interacting case. So long as the background remains dominated by matter or radiation, the analytic solution for each axion field remains same as the single field in Eq.~\eqref{single eom sol}, with $N$ solutions according to the number of axion fields in the model.

In string theory models, axions are expected to couple through a joint potential generated by multiple instantons, and in general the axions will be coupled after making the kinetic terms canonical. The general interacting potential is given by:
\begin{equation}
    V(\phi_1,\dots,\phi_M) = \sum_{i=1}^M \Lambda_i^4 \left[1 - \cos\left(\sum_{j=1}^N \frac{Q_{ij} \phi_j}{f_j} + \delta_i\right)\right],
\end{equation}
where $Q_{ij}$ are charge coefficients and $\delta_i$ are relative phases. We note that the $Q_{ij}$ computed from the instanton charge matrix are in general rational only in a basis where the axion kinetic term is non-diagonal \citep[see e.g.][]{Gendler:2023kjt}. We do not model the kinetic matrix separately, and instead treat the elements of $Q_{ij}$ as continuous random variables, as explained below (i.e. we work in the basis where the kinetic matrix has already been set to the identity).

For two fields, the potential is:
\begin{equation}
    V(\phi_1, \phi_2) = \Lambda_1^4\left(1 - \cos\left(\frac{Q_{11}\phi_1}{f_1} + \frac{Q_{12}\phi_2}{f_2}\right)\right) + \Lambda_2^4\left(1 - \cos\left(\frac{Q_{21}\phi_1}{f_1} + \frac{Q_{22}\phi_2}{f_2}\right)\right),
\end{equation}
where we have set the relative phase to zero using the shift symmetries, and only considered the leading two potential terms that mix the fields. Later we will find it useful to bring the matrix $Q$ into positive upper diagonal form and reabsorb the elements of $Q$ into newly defined decay constants:
\begin{equation}
    V(\phi_1, \phi_2) = m_1^2 \tilde{f}_1^2\left(1 - \cos\left(\frac{\phi_1}{\tilde{f}_1} + \frac{\phi_2}{f_{12}}\right)\right) + m_2^2\tilde{f}_2^2\left(1 - \cos\left( \frac{\phi_2}{\tilde{f}_2}\right)\right).
\end{equation}
We will drop the $\sim$ from $f$ since the context is unambiguous (when we use $f_{12}$, we do not use $Q_{ij}$).

In general, we are allowed to extend this multi-axion potential to $N$-axion fields with positive upper diagonal matrix of $Q$. The potential of the $N$-axion system is given by:
\begin{multline}
    V(\phi_1,\phi_2,\dots,\phi_M)=m_1^2f_1^2\left(1-\cos\left(\frac{\phi_1}{f_1}+\frac{\phi_2}{f_{12}}+\dots+\frac{\phi_M}{f_{1M}}\right)\right)\\+m_2^2f_2^2\left(1-\cos\left(\frac{\phi_2}{f_2}+\frac{\phi_3}{f_{23}}+\dots+\frac{\phi_M}{f_{2M}}\right)\right)+\dots+m_M^2f_M^2\left(1-\cos\left(\frac{\phi_M}{f_M}\right)\right).\label{N-potential}
\end{multline}

In type IIB models such as considered in \cite{Gendler:2023kjt}, the instanton potential is generated by Euclidean D3 branes wrapping effective divisors. The instanton charge matrix can be approximated for the ``prime toric divisors'', of which there are $N+4$ where $N$ is the number of axions. The charge matrix before diagonalising the kinetic term is an $N\times (N+4)$ rectangular integer-valued matrix with an $N\times N$ identitiy sub-block, and thus at least 4 rows giving interactions. The kinetic matrix $K_{ij}$ is ``almost'' diagonal, with eigenvalues spanning an order of magnitude or so: these eigenvalues set the scale of the $f_i$ parameters after diagonalisation. For the $C_4$ and $C_2$ axions, we also have that $K_{ij}$ is independent of the axion fields, as we assumed above. The $\Lambda_i$ in general span an almost scale invariant spectrum, with effective masses extending well below $H_0$. The presence of almost degenerate mass pairs is, however, relatively common.

\subsection{Numerical Method}\label{numerical}

\begin{figure}
  \centering
  \subfloat[Decay constant distribution for single axion model]{\includegraphics[width=.33\linewidth]{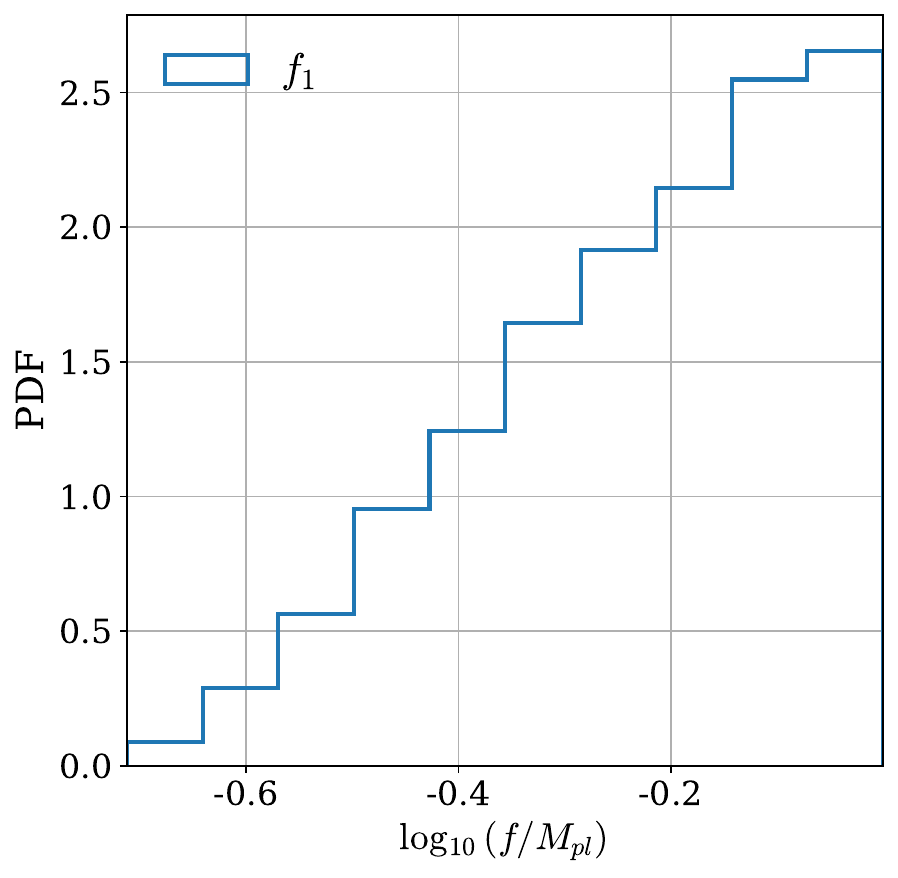}}
  \subfloat[Decay constant distribution for 2 axions without interaction]{\includegraphics[width=.33\linewidth]{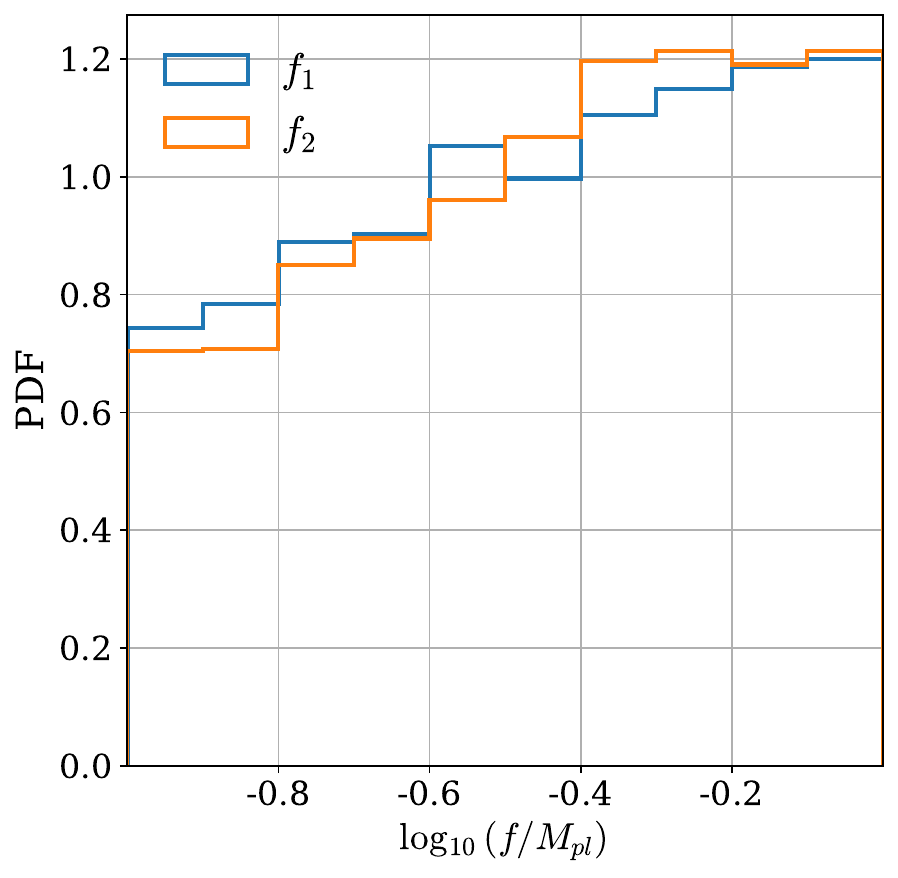}}
  \subfloat[Decay constant distribution for 2 axions with interactions]{\includegraphics[width=.33\linewidth]{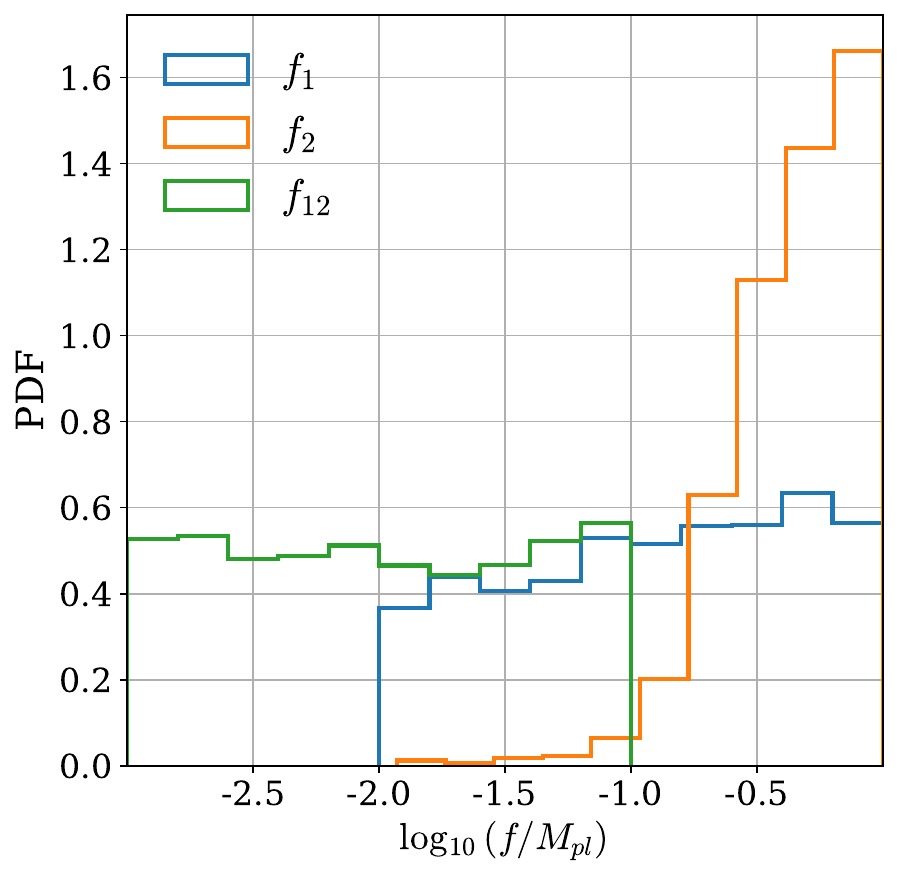}}\\
  \subfloat[Decay constant distribution for 3 axions with interactions]{\includegraphics[width=.33\linewidth]{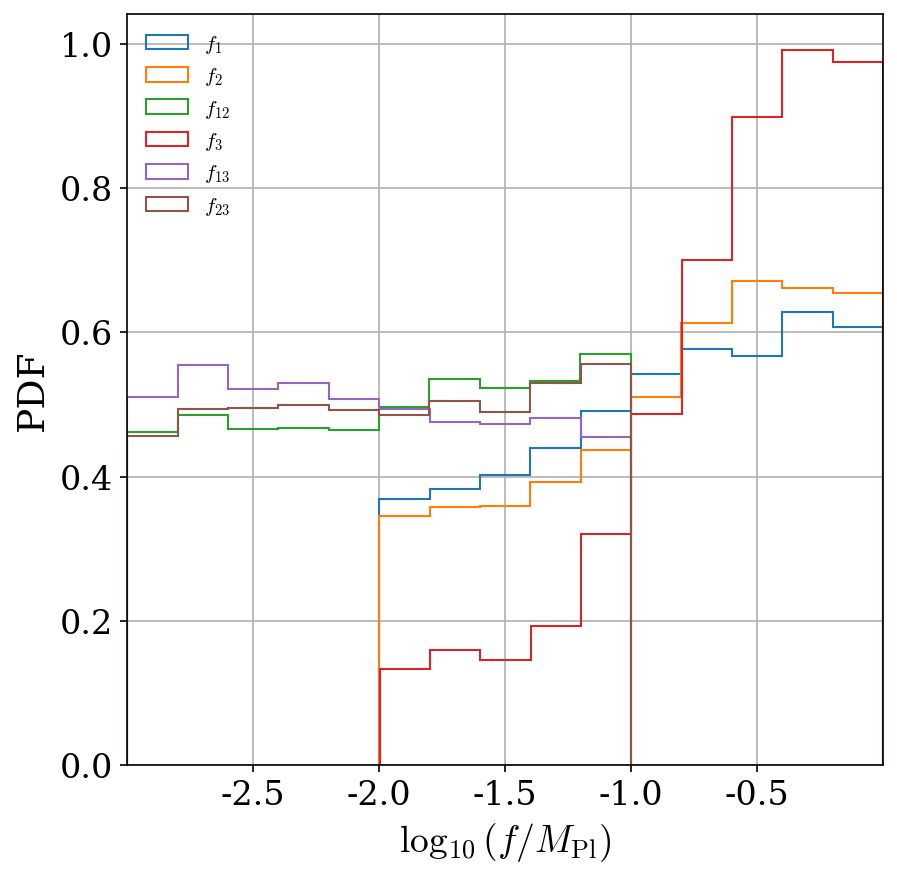}}
  \subfloat[Decay constant distribution for 4 axions with interactions]{\includegraphics[width=.33\linewidth]{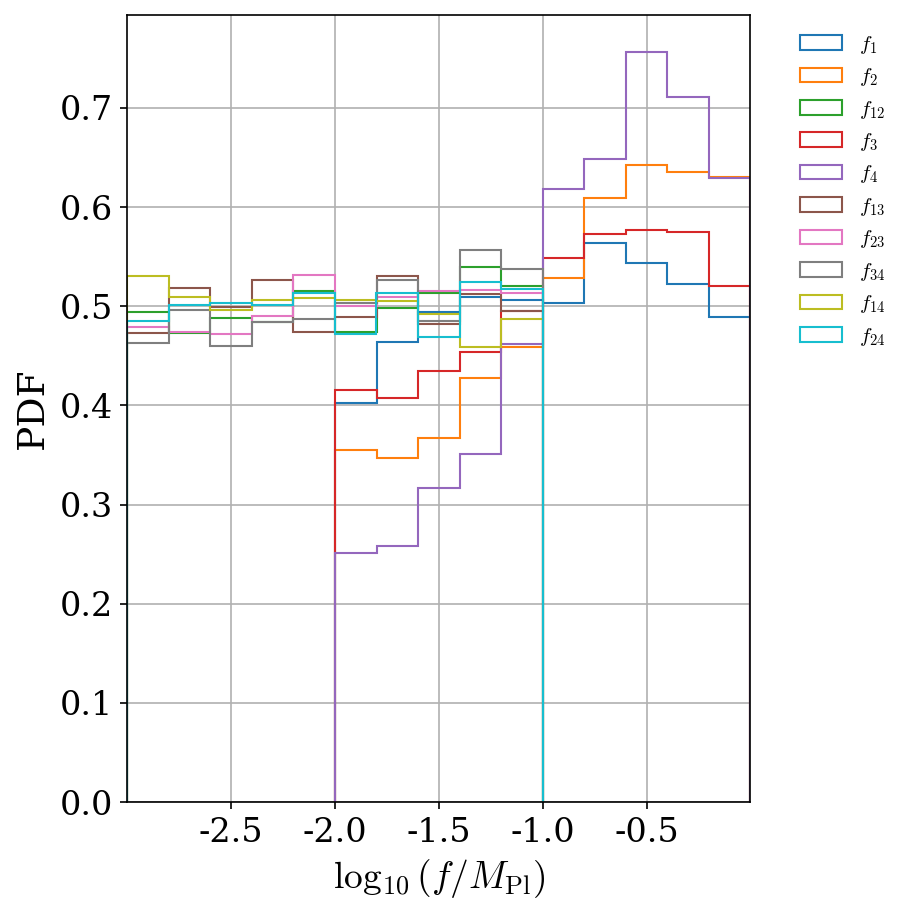}}
  \caption{Decay constant distribution after cuts for each model studied}
  \label{fig:f_hists}
\end{figure}

In this section, we investigate the evolution of the axion field using numerical methods. We solve the axion equations of motion from matter–radiation equality to the late-time universe.

We numerically solve the coupled equations of motion using dimensionless variables:
\begin{equation}
    \tilde{\phi} = \frac{\phi}{M_{\mathrm{Pl}}}, \quad \tilde{t} = m_a t, \quad \tilde{H} = \frac{H}{m_a}.
\end{equation}
By parameterizing our variables, we have the dimensionless equation of motion for a single axion, multiple axion, and an interacting axion, respectively. for the single axion model we have:
\begin{align}
    \ddot{\tilde{\phi}}+3\tilde{H}\dot{\tilde{\phi}}+\frac{f_a}{M_{pl}}\sin(\tilde\phi M_{pl}/f_a)=0
\end{align}
For the two axion non-interacting model we have:
\begin{align}
    \ddot{\tilde{\phi}}_1+3\tilde{H}\dot{\tilde{\phi}}_1+\left(\frac{m_{a_1}}{M}\right)^2\frac{f_{a_1}}{M_{pl}}\sin(\tilde\phi_1 M_{pl}/f_{a_1})&=0\nonumber\\
    \ddot{\tilde{\phi}}_2+3\tilde{H}\dot{\tilde{\phi}}_2+\left(\frac{m_{a_2}}{M}\right)^2\frac{f_{a_2}}{M_{pl}}\sin(\tilde\phi_2 M_{pl}/f_{a_2})&=0
\end{align}
where $M$ is the scaling mass defined as:
\begin{align}
    M(m_{a_1},\dots,m_{a_N})=1/\sum_{i=1}^N\frac{1}{m_{a_i}}
\end{align}
For interacting axions with the charge matrix in upper diagonal form, we have:
\begin{align}
    \ddot{\tilde{\phi}}_1+3H\dot{\tilde{\phi}}_1+\frac{\Lambda_1^4M_{pl}}{f_1M}\sin\left(\frac{\tilde{\phi}_1M_{pl}}{f_1}+\frac{\tilde{\phi}'_2M_{pl}}{f_{12}}\right)=0\nonumber
\end{align}
\begin{align}\label{numer 2a-int}
    \ddot{\tilde{\phi}}_2+3H\dot{\tilde{\phi}}_2+\frac{\Lambda_1^4M_{pl}}{f_{12}M}\sin\left(\frac{\tilde{\phi}_1M_{pl}}{f_1}+\frac{\tilde{\phi}_2M_{pl}}{f_{12}}\right)+\frac{\Lambda_2^4M_{pl}}{f_2M}\sin\left(\frac{\tilde{\phi}_2M_{pl}}{f_2}\right)=0.
\end{align}
The equation of motion of interacting $N$-axion fields is given by:
\begin{gather}
    \ddot{\tilde{\phi}}_1+3H\dot{\tilde{\phi}}_1+\frac{\Lambda_1^4M_{pl}}{f_1M}\sin\left(\frac{\tilde{\phi}_1M_{pl}}{f_1}+\frac{\tilde{\phi}_2M_{pl}}{f_{12}}+\dots+\frac{\tilde{\phi}_M M_{pl}}{f_{1M}}\right)=0,\nonumber\\
    \ddot{\tilde{\phi}}_1+3H\dot{\tilde{\phi}}_1+\frac{\Lambda_1^4M_{pl}}{f_1M}\sin\left(\frac{\tilde{\phi}_1M_{pl}}{f_1}+\frac{\tilde{\phi}_2M_{pl}}{f_{12}}+\dots+\frac{\tilde{\phi}_M M_{pl}}{f_{1M}}\right)+\frac{\Lambda_2^4M_{pl}}{f_2M}\sin\left(\frac{\tilde{\phi}_2M_{pl}}{f_{2}}+\dots+\frac{\tilde{\phi}_M M_{pl}}{f_{2M}}\right)=0,\nonumber\\
    \vdots
\end{gather}
Thus, our general Friedmann's equation becomes
\begin{align}
    \left(\frac{\tilde{\dot a}}{a}\right)^2= \frac{1}{3} \left\{(\tilde{\rho}_{r,0}/a^{-4} + \tilde{\rho}_{m,0}/a^{-3}) 
+ \sum_{i=1}^N\frac{1}{2} \dot{\tilde{\phi}}_i^2 
+ \frac{V\left(\tilde{\phi}_1,\tilde{\phi}_2,\dots\tilde{\phi}_M\right)}{M^2M_{pl}^2}\right\}
\end{align}
where $\tilde{\rho}_{r,0}$ and $\tilde{\rho}_{m,0}$ are the dimensionless energy densities of radiation and matter, respectively. We parametrize the energy densities as $\tilde{\rho}_{r,m} = \frac{\rho_{r,m}}{M^2 M_{\mathrm{pl}}^2},$ and we set the present--day values to $\rho_{r,0} = 2.1 \times 10^{-47}\,\mathrm{GeV}^4$ and $\rho_{m,0} = 7.9 \times 10^{-51}\,\mathrm{GeV}^4$.

We evolve the system from the early radiation--dominated era to the present day, ensuring a correct matching of $\Omega_m$ and $\Omega_r$ to the \emph{Planck} parameters~\cite{Planck:2018vyg}. The axion parameters are varied to investigate the influence of mass hierarchy, initial misalignment angles, and field couplings.

\begin{figure}
    \centering
    \includegraphics[width=0.9\linewidth]{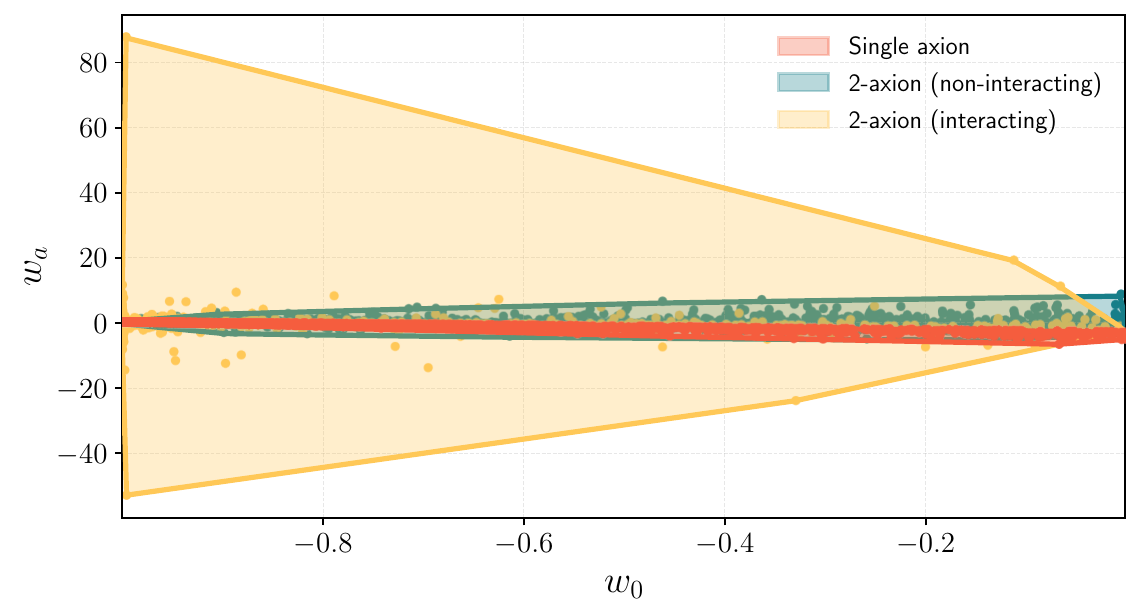}
    \caption{Equation of state distribution. We show scattered points from our Monte Carlo model drawn from the prior and post selected on $\Omega_\phi$ and $h$. We further focus on the region with $w_0<0$, and delineate the convex hull in the parameter space: i.e. the outermost regions explored by the priors. The two-field and two-field interacting models explore regions far outside the thawing quintessence region given by the single field model.}
    \label{eos}
\end{figure}

\begin{table}
    \centering
    \begin{tabular}{|c|c|c|}
    \hline
    Model&Parameter&Sampled Values\\
    \hline
      Single Axion Quintessence   & $\phi_i$ & $f_a/M_{pl}\times\mathcal{U}[0,\pi]$ \\
        & $\log_{10}m_a$ & $\mathcal{U}[-33,-32]$\\
        & $\log_{10}\left(\frac{f_a}{M_{pl}}\right)$&$\mathcal{U}[-1,0]$\\
        \hline
        Non-interacting 2-axion Quintessence& $\phi_{i1},\phi_{i2}$ & $f_{1,2}/M_{pl}\times\mathcal{U}[0,\pi]$ \\
        & $\log_{10}m_{a_1},\log_{10}m_{a_2}$ & $\mathcal{U}[-33,-32]$\\
        & $\log_{10}\left(\frac{f_{a_1}}{M_{pl}}\right),\log_{10}\left(\frac{f_{a_2}}{M_{pl}}\right)$&$\mathcal{U}[-1,0]$\\
        \hline
        Interacting 2-axion Quintessence& $\phi_{i1},\phi_{i2}$ & $f_{1,2}/M_{pl}\times\mathcal{U}[0,\pi]$ \\
        & $\log_{10}m_{a_1},\log_{10}m_{a_2}$ & $\mathcal{U}[-33,-32]$\\
        & $\log_{10}\left(\frac{f_{a_1}}{M_{pl}}\right),\log_{10}\left(\frac{f_{a_2}}{M_{pl}}\right)$&$\mathcal{U}[-2,0]$\\
        & $\log_{10}\left(\frac{f_{12}}{M_{pl}}\right)$&$\mathcal{U}[-3,-1]$\\
        \hline
    \end{tabular} 
    \caption{Parameters and priors. For the 3 and 4-axion models we only consider the interacting case and take generalised priors the same as the 2-axion model.}
    \label{initial}
\end{table}
By setting the parameter value ranges in Table~\ref{initial}, we assigned a uniform probability distribution to each model parameter using the \texttt{numpy.random.uniform} function, and using \texttt{scipy.integrate.odeint} to integrate our differential equations. For every simulation, the initial values of the axion fields were sampled in the range $0$ to $\pi$. This simple sampling method for the initial values $\phi_i$ gives a uniform grid in the original field basis, but is not uniform in the fundamental domain defined by the potential \citep[see][and references therein for related discussions also counting the number of distinct vacua]{Gendler:2023hwg}. The axion masses were sampled between $10^{-33}\,\mathrm{eV}$ and $10^{-32}\,\mathrm{eV}$, corresponding to the ultralight scalar field regime commonly considered for quintessence.
For the decay constants, we set the distributions in both the single-axion and non-interacting axion quintessence models to lie close to the Planck scale. In the interacting axion quintessence case, the decay constants $f_1$ and $f_2$ were similarly chosen to be near the Planck scale, while the coupling decay constant $f_{12}$ was sampled such that $\log_{10}(f_{12}/M_{\mathrm{pl}}) \in [-3,-1]$. This choice ensures that the interaction strength between the two axion fields remains significant but remains below the Planck-scale suppression.\\
\indent To set the initial time for the numerical evolution, we define the dimensionless initial time as
\[
\tilde{t}_i = \frac{A^2}{2\,\tilde{H}_{\mathrm{eq}}},
\]
where $A=10^{-2}$, and $\tilde{H}_{\mathrm{eq}}$ denotes the Hubble parameter at matter-radiation equality. In our dimensionless units, it is expressed as
\[
\tilde{H}_{\mathrm{eq}} = \sqrt{2\,\tilde{\rho}_{m,0}(1+z_{\mathrm{eq}})^3},
\]
with $z_{\mathrm{eq}} = 3400$ the redshift at equality. This choice ensures that the evolution begins sufficiently early in the radiation--dominated era, allowing the axion field to evolve naturally from its frozen misalignment state.

For the scale factor, we set the initial condition as $a_i = \frac{A}{1 + z_{\mathrm{eq}}},$
which guarantees a self-consistent normalization of the background expansion history. Starting the integration thus before $z_{\mathrm{eq}}$ allows the radiation and matter components to be correctly weighted according to
$\rho_r(a_i)$ and $\rho_m(a_i),$. With this setup, the axion dynamics is evolved from the radiation era through matter domination to the present day, which is defined to be when $a=1$.
.\\

\section{Results and Discussion}\label{results}
\subsection{Axion Dynamics}\label{sec:dynamics}

\begin{figure}
  \centering
  \subfloat[$m_1=1.27\times10^{-42}\text{GeV},\ m_2=3.17\times10^{-42}\text{GeV},\ f_1/M_{pl}\approx 0.129,\ f_2/M_{pl}\approx0.0233,\ f_{12}/M_{pl}\approx 0.0202,\ \tilde{\phi}_{1i}\approx0.297,\ \tilde{\phi}_{2i}\approx0.006,\
h=0.738,\ \Omega_\phi=0.770,\ w_\phi\approx-0.811,\ w_a\approx-0.390$]{\includegraphics[width=0.9\linewidth]{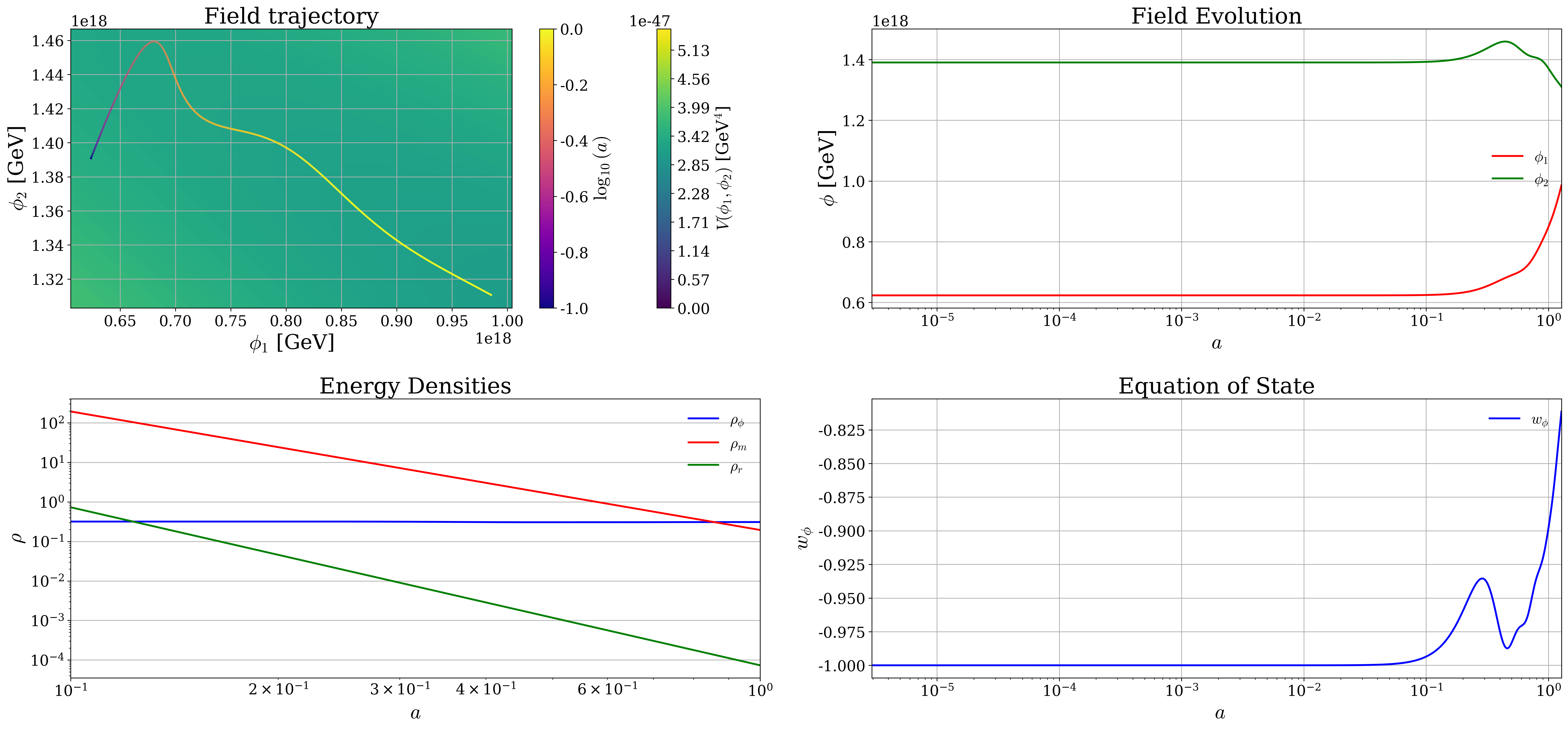}}\par\medskip
  \subfloat[$m_1=1.50
\times10^{-42}\text{GeV},\ m_2=4.48
\times10^{-42}\text{GeV},\ f_1/M_{pl}\approx 0.046,\ f_2/M_{pl}\approx0.195,\ f_{12}/M_{pl}\approx 0.0016,\ \tilde{\phi}_{1i}\approx0.0047,\ \tilde{\phi}_{2i}\approx0.3507,\ h=0.622,\ \Omega_\phi=0.673,\ w_\phi\approx-0.897,\ w_a\approx-12.4$]{\includegraphics[width=0.9\linewidth]{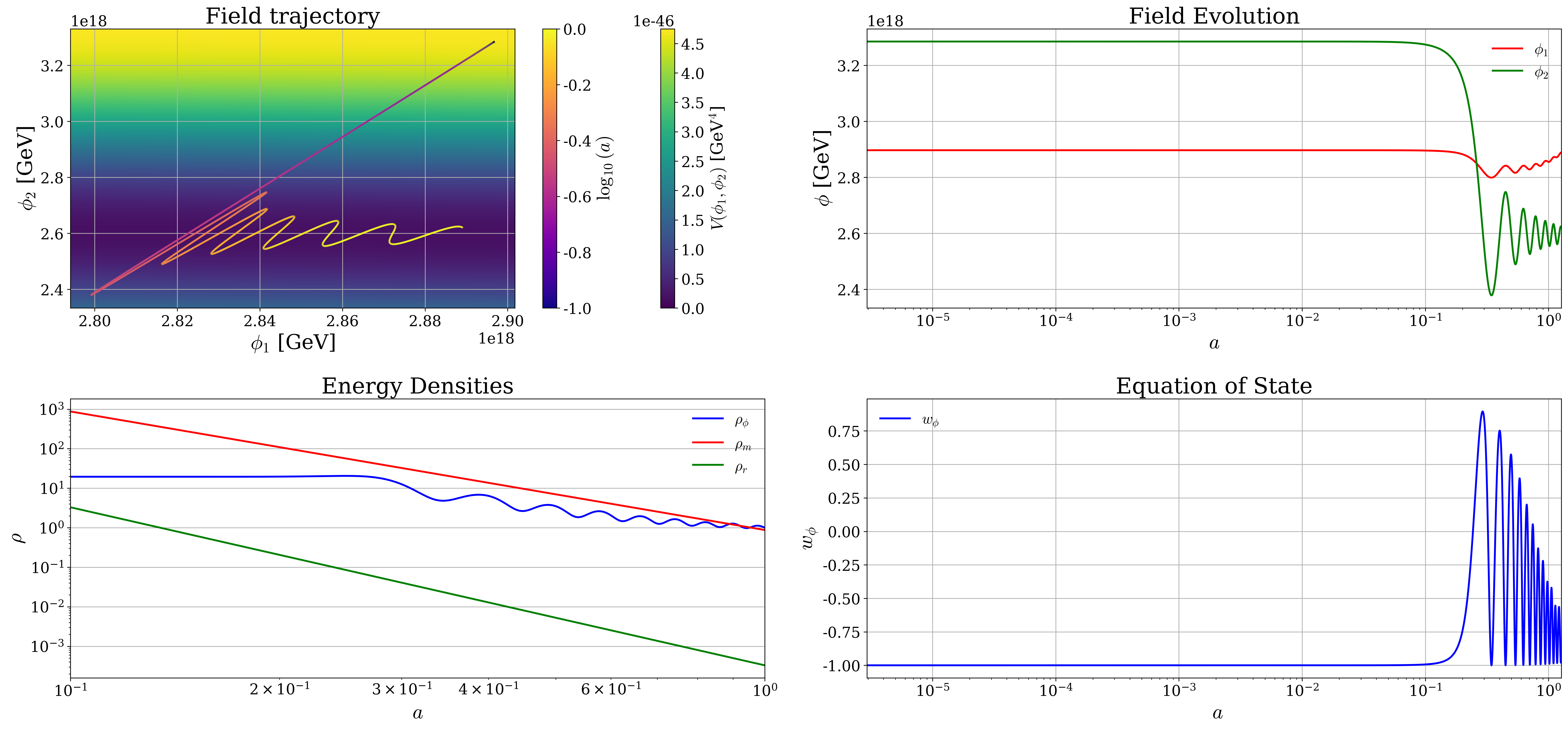}}
  \caption{Two axion evolution (1)}
\label{evo1}
\end{figure}

To study the axion DE parameters, we define the derivative of equation of state as $w'=\frac{d\omega}{d\ln a}$, which when evaluated at $a=1$ gives $w'=-w_a$ in the CPL parameterisation. By setting the initial values of our parameters as TABLE I in each model and using the time interval from matter-radiation equality to the late-time universe, we evaluate the joint distirbution of $w_0,w_a,\Omega_\phi$ and $h$. For each model we generate generate $8\times10^4$ samples following the priors outlined in the previous section, and filter the output data to select only those models yielding $0.6<\Omega_\phi<0.8$ and $0.6<h<0.8$, roughly aligning to current inferences, and suitable as priors for cosmological analysis such as \cite{Urena-Lopez:2025rad}. Defining the final time to be at $a=1$ fixes the matter and radiation densities based on the density at equality, but leaves the Hubble parameter, relative fraction of ordinary matter to DE, and the age of the Universe as derived parameters. For the single field model and the interacting two field model, approximately 4\% of the Monte Carlo samples satisfy the cuts on on $\Omega_\phi$ and $h$, while for the non-interacting two field model this fraction rises to around 10\%, indicating that it is easier to achieve large energy densities without the added dynamical complexity introduced by interactions.

Fig.~\ref{fig:f_hists} shows the normalised probability density function (PDF) for the decay constant parameters after cuts for each model. We see that the single field model has, as is well known, a strong preference for $f_a\sim M_{pl}$, while the two field non-interacting model has a flatter distribution allowing smaller decay constants. For the interacting two field model, one decay constant, $f_2$, follows a distribution close to the single field one, while the other two decay constants, $f_1$ and $f_{12}$ show log-flat distributions consistent with the priors. For three and four axions model, it can be seen that the density of diagonal element decay constants $(f_1,f_2,f_3 ,\,\text{and}\; f_4)$ also peaks at Planckian values as we increase the number of axion fields. This further illustrates that the model we have for interactions actually hinders DE behaviour compared to a non-interacting model. As we now show, however, the presence of interactions among cosines allows for significantly more exotic behaviour in the dynamics and equation of state.

Fig.~\ref{eos} compares the two field models by their DE equation of state parameters. Points show the scatter in these parameters based on the prior after cuts, restricted to $w_0<0$ for visualisation. The lines denote the convex hull given by the parameters, i.e. the outer most region explored. Fig.~\ref{eos} shows clearly how the single field model lies in the ``thawing quintessence'' region with $w_0>-1$ and $w_a<0$ preferred by DESI. However, in models with two fields the value of $w_a$ can be positive, which is not possible for the single field model. Furthermore, the magnitude of $w_a$ varies far more widely in the interacting than the non-interacting case, yielding extreme models. We observed the same effect in the three and four field models that the prior gains support over a wider range of values.
\begin{figure}
  \centering
  \subfloat[$m_1=2.51
\times10^{-42}\text{GeV},\ m_2=1.93
\times10^{-42}\text{GeV},\ f_1/M_{pl}\approx 0.847,\ f_2/M_{pl}\approx0.0104,\ f_{12}/M_{pl}\approx 0.0029,\ \tilde{\phi}_{1i}\approx2.5405,\ \tilde{\phi}_{2i}\approx0.03136,h=0.738,\ \Omega_\phi=0.770,\ w_\phi\approx-0.811,\ w_a\approx-0.390$]{\includegraphics[width=0.9\linewidth]{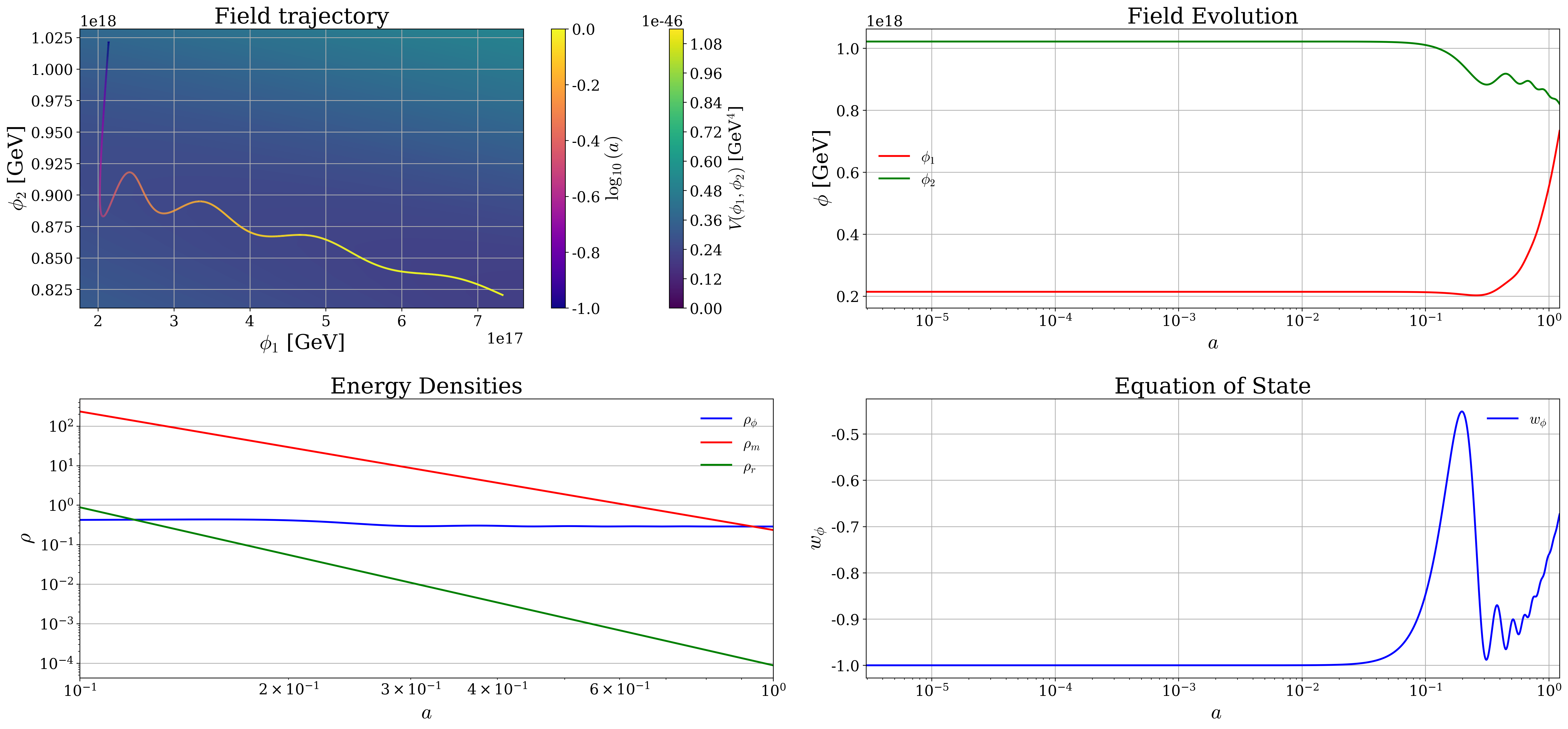}}\par\medskip
  \subfloat[$m_1=6.26
\times10^{-42}\text{GeV},\ m_2=5.98
\times10^{-42}\text{GeV},\ f_1/M_{pl}\approx 0.940,\ f_2/M_{pl}\approx0.1562,\ f_{12}/M_{pl}\approx 0.0027,\ \tilde{\phi}_{1i}\approx0.589,\ \tilde{\phi}_{2i}\approx0.266, h=0.688,\ \Omega_\phi=0.720,\ w_\phi\approx-0.939,\ w_a\approx 1.68$]{\includegraphics[width=0.9\linewidth]{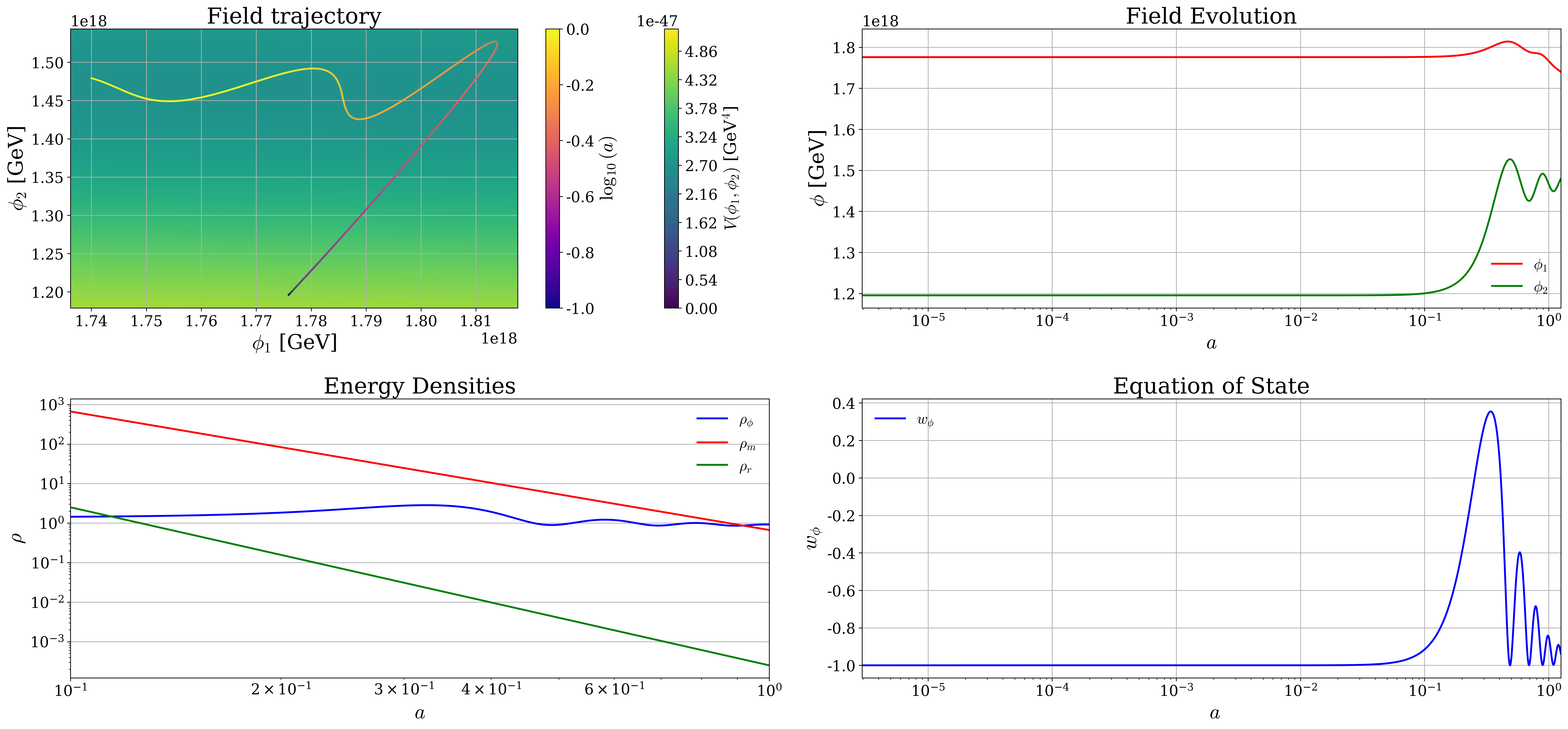}}
  \caption{Two axion evolution (2)}
\label{evo2}
\end{figure}

We now investigate the dynamics of certain special cases of the two-field model in detail in Figs.~\ref{evo1}, \ref{evo2}. These two figures show the dynamics of axion fields $\phi_1$ and $\phi_2$ in in the plane $(\phi_1,\phi_2)$ in the upper left panel and $\phi_{1,2}(a)$ in the upper right panel. In the bottom left panel we show the density, $\rho_i(a)$ of the total DE compared to matter and radiation, and in the lower right panel the total DE equation of state $w_\phi(a)$. We only show dynamics for the interacting models. In the non-interacting case the fields evolve independently since our initial conditions have no angular momentum, and oscillations further decouple evolution from the background.

In Figure \ref{evo1}, panel (a) shows the field rolling slowly along the potential without strong oscillations. Both $\phi_1$ and $\phi_2$ evolve smoothly with only small variations, and the density plot remains nearly flat until late times, then tilts downward, corresponding to a moderately evolving equation of state with $w_\phi \approx -0.81$ and $w_a \approx -0.39$. Panel (b) illustrates sharper behavior, with the field encountering a steep feature in the potential, executing pronounced oscillations in $\phi_1$ and $\phi_2$, which imprint as rapid wiggles in the energy density $\rho_\phi$, signaling a transition toward matter-like scaling. The equation of state here becomes extremely positive, $w_a \approx 12$, indicating very rapid recent evolution. This evolution displays, however, the very interesting feature that although the fields are rapidly oscillating, the equation of state is not itself oscillating around zero, but around an increasingly small value which at the end has $\langle w\rangle\approx -0.75$. In the Appendix we show how such beahviour can be understood in terms of a rotating eigenbasis model induced by axion interactions. Oscillation approaching $w=-1$ can be explained by a false vacuum.

Figure \ref{evo2}, panel (a) shows an intermediate scenario with gentle oscillations in the components and smooth, smaller fluctuations in the density. The equation of state again displays the feature of being oscillatory, but remaining negative with a strong change at late times caused by a turn in the field trajectory as the second field starts to evolve. In Figure \ref{evo2}, panel (b) shows the field crossing an inflection-like region of the potential, with the components evolving with even milder oscillations.  The equation of state is $w_\phi \approx -0.94$ and $w_a \approx 1.68$, suggesting that the field is slowly becoming less negative in its equation of state. However due to the rapid oscillations the instantaneous value of $w_a$ is misleading, and we see that the average $\langle w_\phi\rangle$ is in fact becoming more negative, and again demonstrating the oscillating-but-negative feature. 

\begin{figure}
    \centering
    \includegraphics[width=0.9\linewidth]{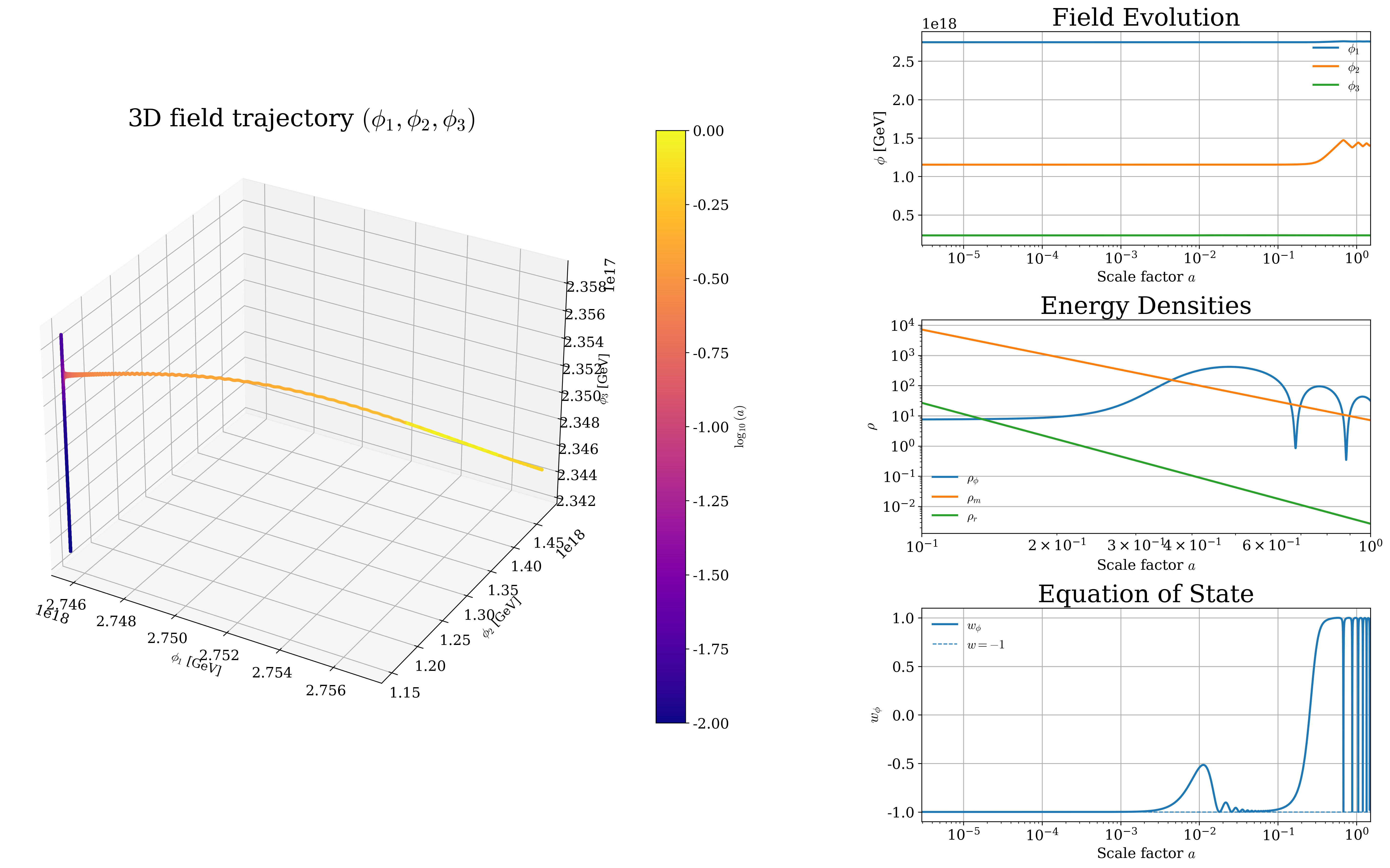}
    \includegraphics[width=0.9\linewidth]{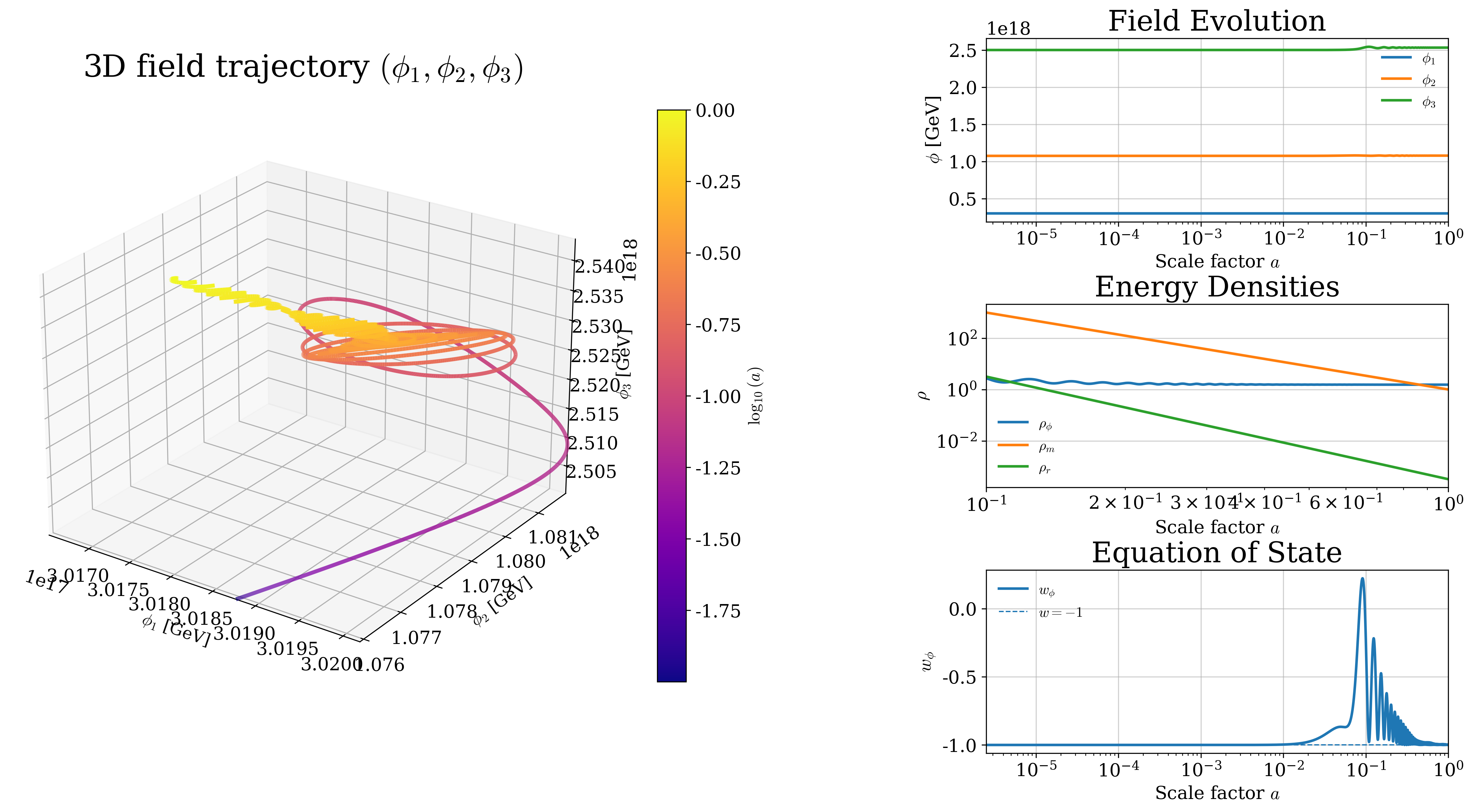}
    \caption{Three axion evolution (1). Top: $w_\phi=-0.95$, $w_a=1.36$, $\Omega_{\phi,0}=0.67$, and $h_0=0.66$. Bottom: $w_\phi=-1.00$, $w_a=4.24$, $\Omega_{\phi,0}=0.61$, and $h_0=0.63$.}
    \label{fig:3a1}
\end{figure}

\begin{figure}
    \centering
    \includegraphics[width=0.9\linewidth]{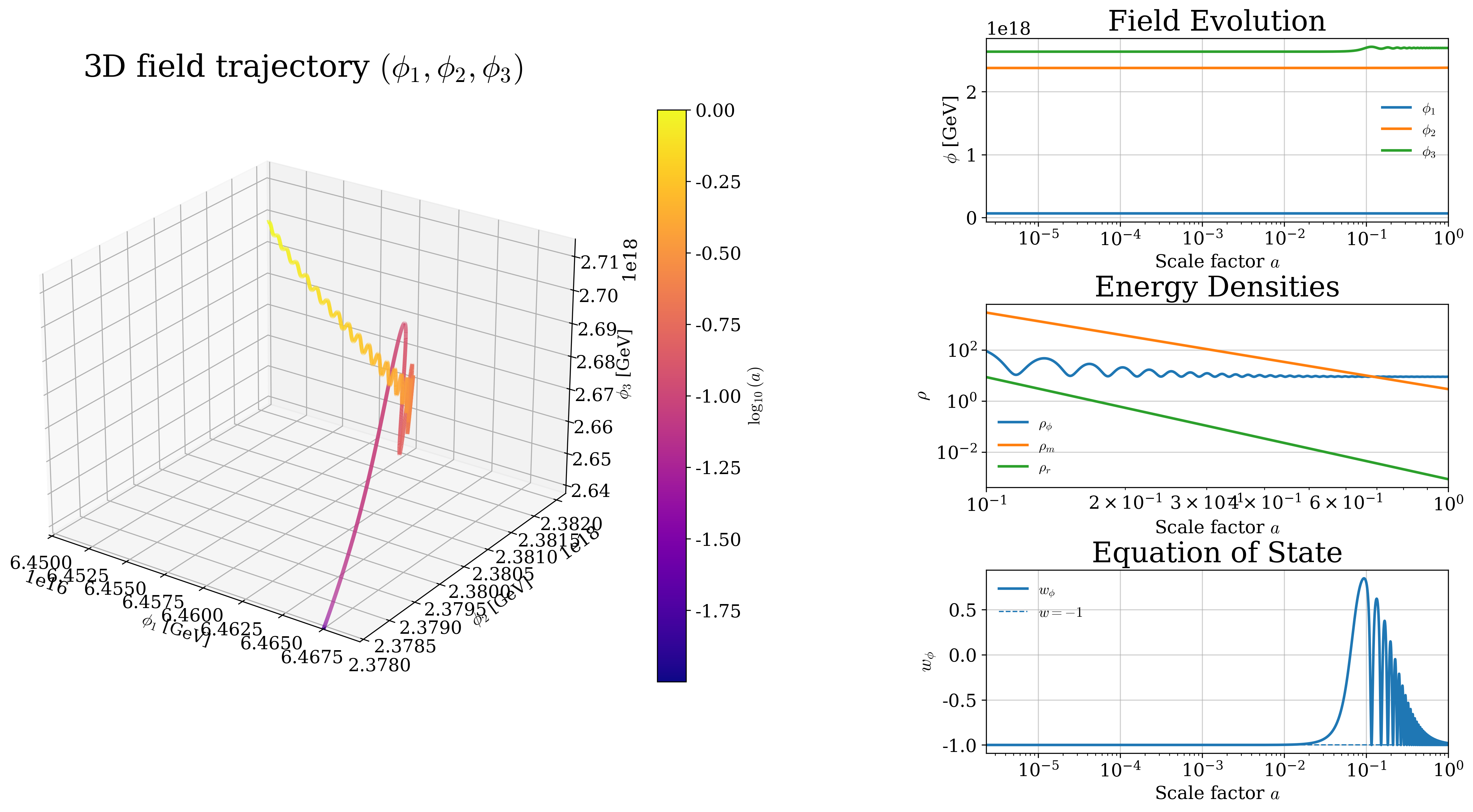}
    \includegraphics[width=0.9\linewidth]{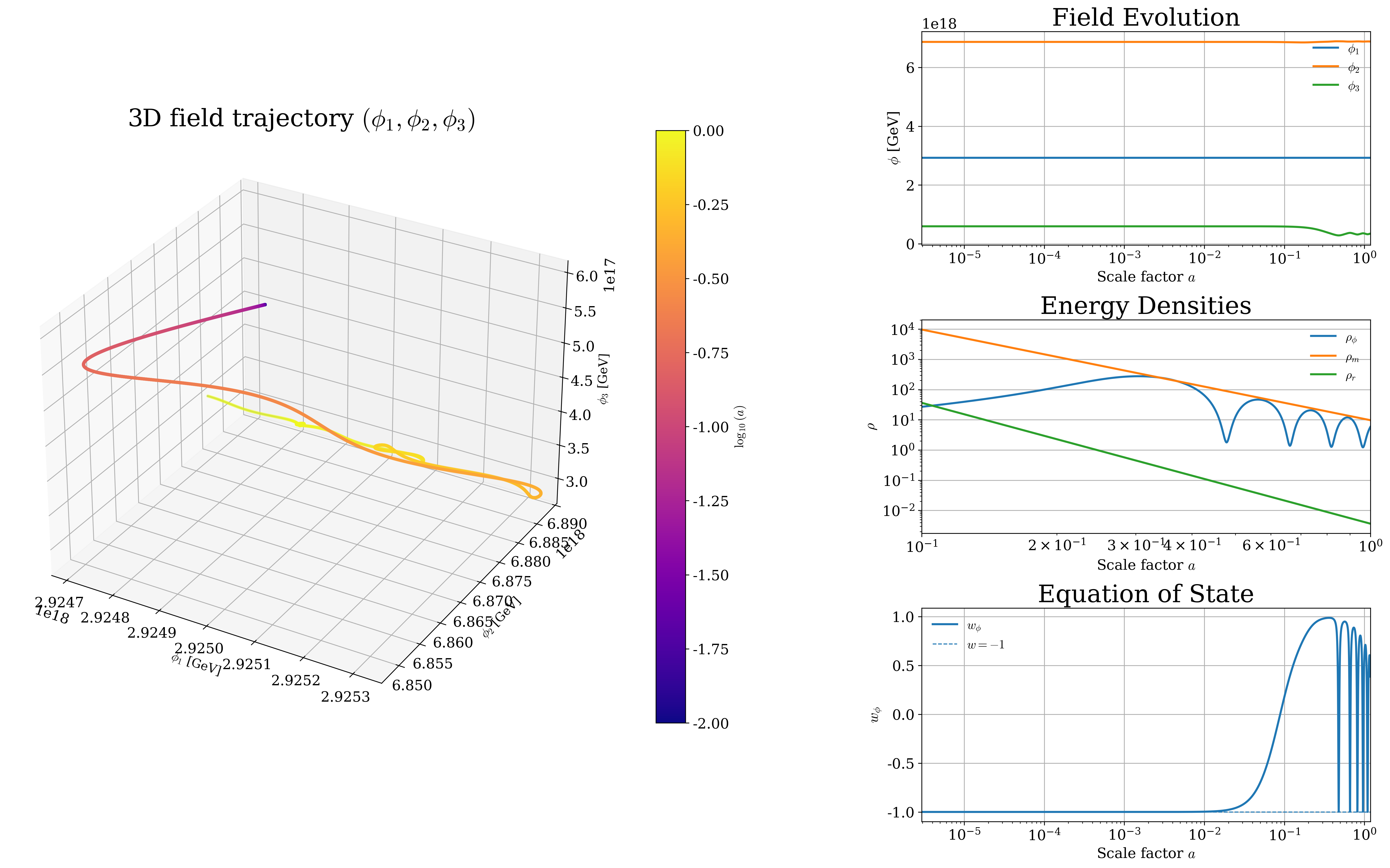}
    \caption{Three axion evolution (2). Top: $w_\phi=-0.98$, $w_a=-1.48$, $\Omega_{\phi,0}=0.75$, and $h_0=0.72$. Bottom: $w_\phi=-0.36$, $w_a=-28.7$, $\Omega_{\phi,0}=0.67$, and $h_0=0.61$.}
    \label{fig:3a2}
\end{figure}

\begin{figure}
    \centering
    \includegraphics[width=\linewidth]{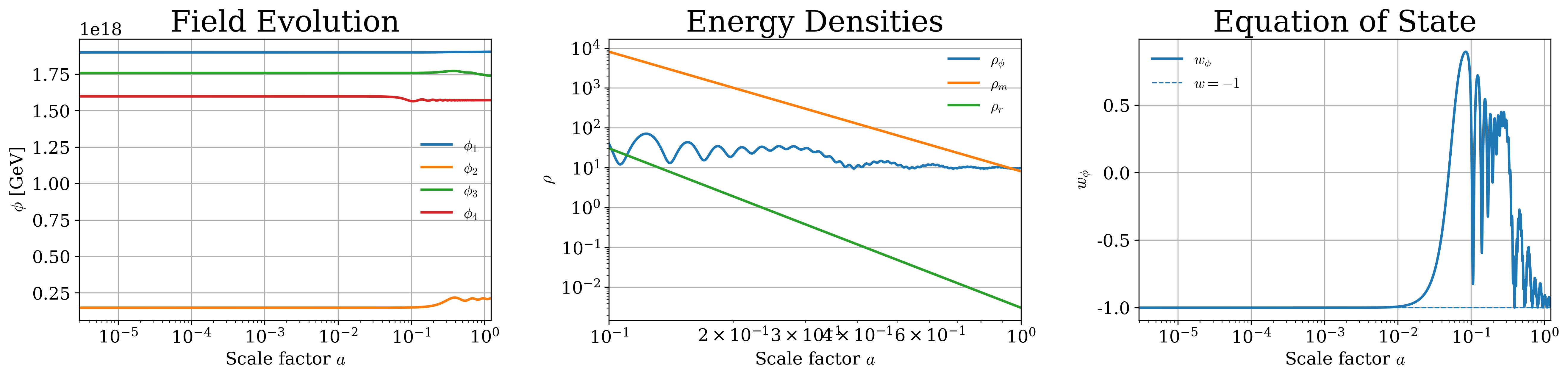}
    \includegraphics[width=\linewidth]{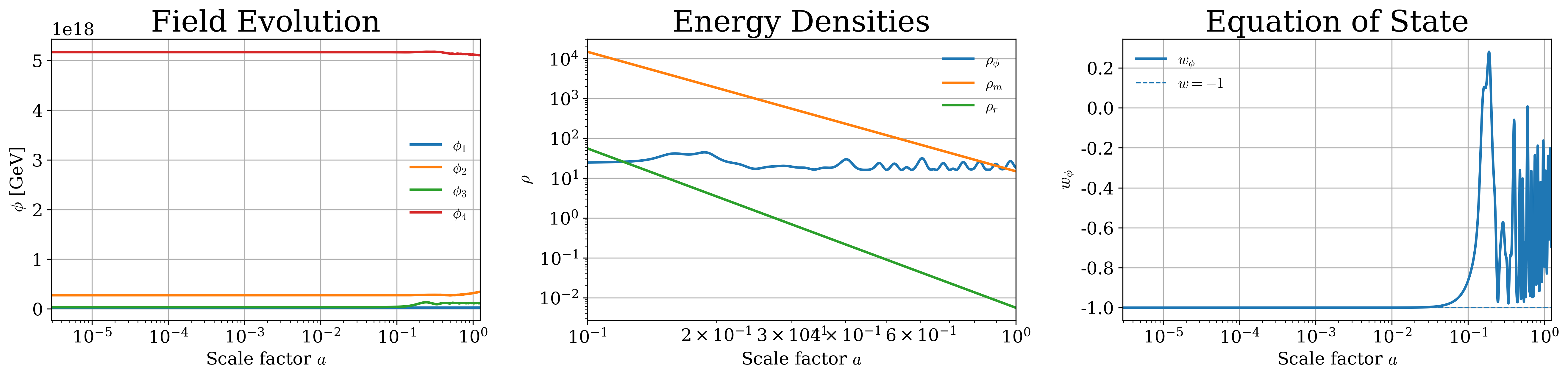}
    \includegraphics[width=\linewidth]{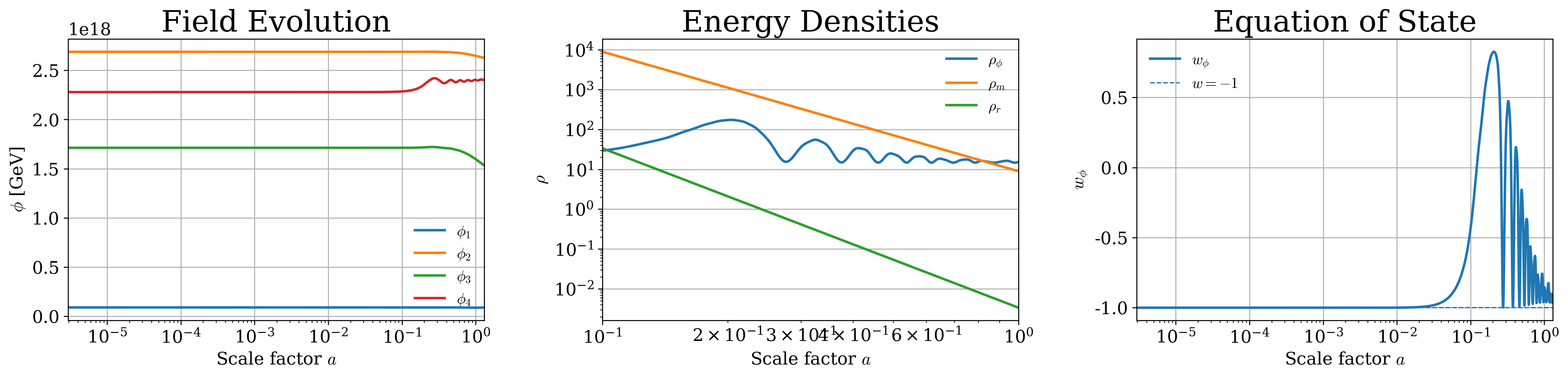}
    \caption{Four axion evolution. Top: $w_\phi=-0.98,\; w_a=-2.17,\; \Omega_{\phi,0}=0.68,\; h_0=0.67$. Middle: $w_\phi=-0.90,\; w_a=-0.71,\; \Omega_{\phi,0}=0.79,\; h_0=0.74$. Bottom: $w_\phi=-0.90,\; w_a=-0.71,\; \Omega_{\phi,0}=0.79,\; h_0=0.74$.}
    \label{fig:4a}
\end{figure}

In Figs.~\ref{fig:3a1}, \ref{fig:3a2} we present the evolution of three axion systems, showing 3D field trajectories $(\phi_1,\phi_2,\phi_3)$ in the left panel, and field evolution $(\phi-t)$, energy densities evolution $(\rho-a)$, and equation of state evolution in the right panels. Fig.~\ref{fig:4a} illustrates the field evolution of the four axion quintessence model. In all the multifield models we find similar behaviours to the two field case, but more extreme variation in particular with $w$ including multiple oscillations scales, and apparently chaotic trajectories. We also observe the rotation of the mass eigenbasis, as in the two-field case. However, it has a more complicated structure because the $N$-field system contains more than one independent mixing direction. In contrast to the two-field case, where the rotation can be described by a single mixing angle, the rotation in the multi-field system requires a general $SO(N)$ transformation as shown in Appendix \ref{analy-N}. 
\begin{figure}
    \centering
    \includegraphics[width=\linewidth]{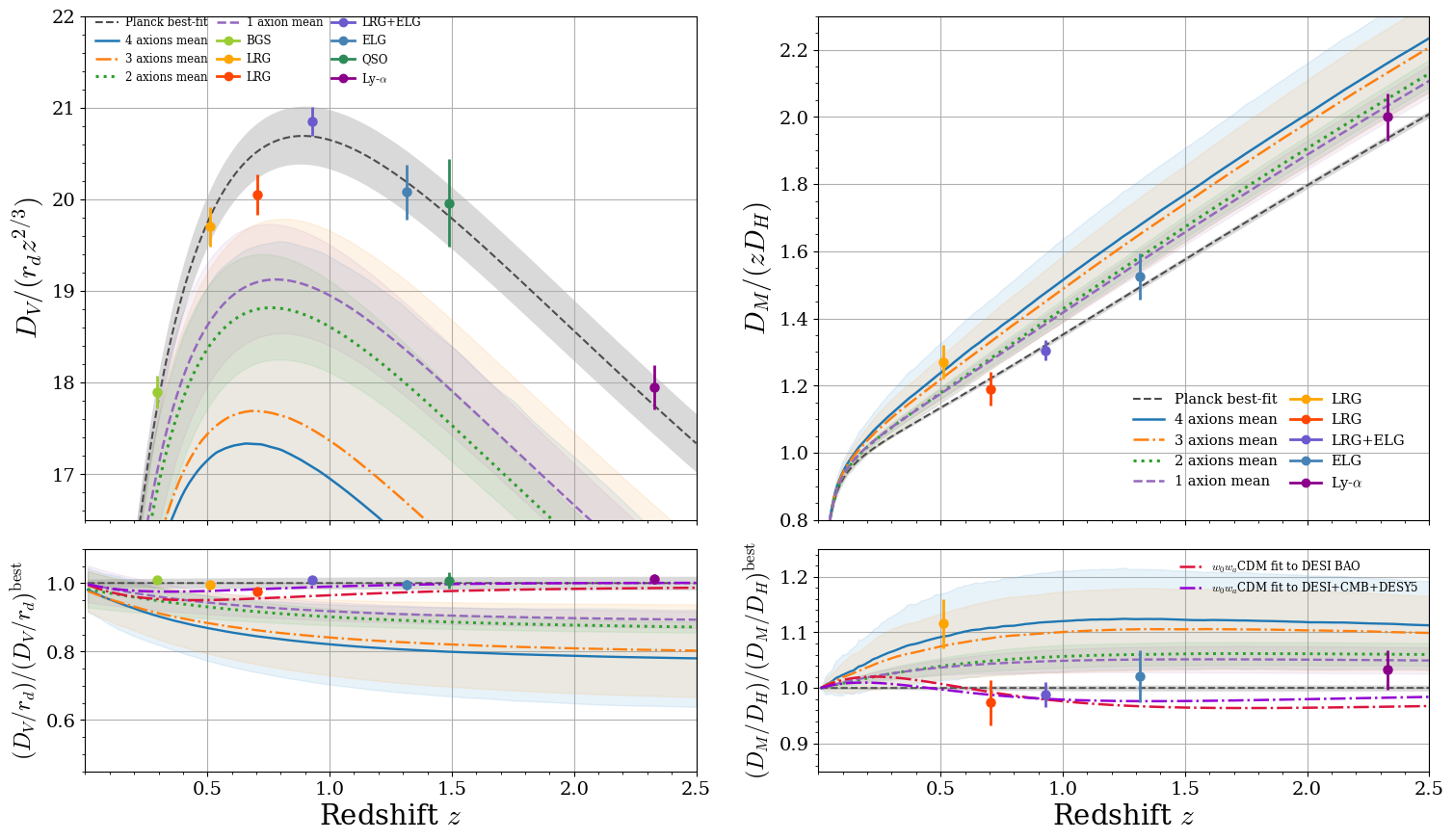}
    \caption{Top left:  the ratio of the angle-averaged distance $D_V\equiv(z\ D_M^2D_H)^{1/3}$ to the sound
horizon at the baryon drag epoch, $r_d$. Top right: the ratio of transverse and line-of-sight comoving
distances $D_M/D_H$, and redshift $z$. Bottom left: the ration of $D_V/r_d$ of each model and $D_V/r_d$ of $\Lambda CDM$ best fit. Bottom right: the ratio of $D_M/D_H$ of each model and $D_M/D_H$ of to the Planck best fit. Datasets and labeling are as in \citep{DESI:2024mwx}.}
    \label{fig:distance-plot}
\end{figure}

\subsection{Distance Measures}
We now consider the BAO distance measures from DESI \citep{DESI:2024mwx}. There are three relevant quantities: the transverse comoving distance $D_M$, the Hubble distance $D_H$, and the angle-average distance, $D_V$. For simplicity, we consider only flat FRW models where we have:
\begin{align}
    D_M^{\text{flat}}(z)&=\int_0^z\frac{cdz'}{H(z')},\\
    D_H^\text{flat}(z)&=\frac{c}{H(z)},\\
    D_V^\text{flat}(z)&=\left[ z\ D_M^2(z) D_H(z) \right]^{1/3}.
\end{align}
Moreover, BAO measurements depend on the sound horizon at the drag epoch, which is given by:
\begin{align}
    r_d=\int_{z_d}^{\infty}\frac{c_s(z')}{H_d(z')} \, dz',
\end{align}
where $c_s(z)$ is the sound speed of the tightly coupled photon-baryon fluid prior to recombination,
\begin{align}
    c_s(z)=\frac{1}{\sqrt{3\left(1+\frac{3\rho_B(z)}{4\rho_\gamma(z)}\right)}}\,,
\end{align}
and $\rho_B(z)$ and $\rho_\gamma(z)$ are the baryon and photon energy densities, respectively (to fix $\rho_\gamma$ from the value of $\rho_r$ specified in Section~\ref{Theory} we assume three massless neutrinos). Since $r_d$ is determined by early-Universe physics, the integral is dominated by redshifts around the drag epoch, $z_d\approx 1060$, which is close to the recombination era. Since radiation and matter are the main components in the early-time universe (the quintessence density is assumed very small compared to other components, i.e. our models are chosen to not give any early DE), we approximate the Hubble parameter at early times as:
\begin{align}
    H_d^2(z') = \frac{1}{3M^2_{pl}}\left[\frac{\rho_{m,0}}{a^{3}}+\frac{\rho_{r,0}}{a^4}\right].
\end{align}
In contrast, the Hubble parameter for the comoving distance calculation is given by our numerical solution including axions, i.e. Eq.~\eqref{eqn:full_friedmann}.\\

Fig. \ref{fig:distance-plot} illustrates the comparison of the distance-redshift relation between the Planck best-fit curve and its $1\sigma$ uncertainty band \citep{Planck:2018vyg}, and the prior mean and $1\sigma$ of the interacting multi-axion quintessence models. The results show that increasing the number of axion fields produces a larger deviation from the standard $\Lambda$CDM and Planck distance relation, especially at higher redshift. This behavior can be seen more clearly from the ratio plots in the lower panels, where the multi-axion models gradually move away from unity as $z$ increases. For the volume-averaged BAO distance, $D_V/(r_d z^{2/3})$, the interacting multi-axion models predict smaller distances than the best-fit $\Lambda$CDM curve, with the $N=4$ case giving the strongest suppression, followed by $N=3$ and $N=2$. In contrast, for the transverse comoving distance quantity, $D_M/(zD_H)$, the multi-axion models give larger values than the $\Lambda$CDM best fit, and the enhancement again becomes stronger as the number of axion fields increases. Furthermore, by comparing with $w_0w_a$CDM in the residuals we see how, as expected, the single axion model can most closely mimic this only at low redshift, especially in $D_V$.

Exotic behaviour of distance meausres in example models are shown in Appendix~\ref{sec:C}. Strong oscillations in the axion quintessence energy density can be passed on to the distance measures. If such exotic beheviour were ever observed at high significance in future precision data, this could favour multi-axion quintessence. Such a possibility, however intriguing, nonetheless seems unlikely.

\section{Conclusions}\label{sec:conclusions}

We have studied multi-field axion quintessence models with two, three or four fields, sampling the parameters according to basic principles. We conditioned the fields to have small masses near $H_0$ and large decay constants near $M_{pl}$, and allowed for an off-diagonal interaction in the cosine potential. We sampled the dimensionful parameters with log-uniform priors, and the initial conditions on the angle with a linear uniform prior. We conditioned the output of our numerical solution to have present day Hubble parameter and quintessence energy density in a broadly acceptable prior range. We found that the interaction term in the joint potential can have interesting effects.

The interacting model requires, statistically, at least one decay constant as large as the single field model. The presence of interactions seems to hinder the ability of multiple fields to work together in an ``N-flation'' like manner that is possible with two non-interacting fields. This is likely due to stronger non-linearities.

The interacting model displays much more diversity in the behaviour of the equation of state compared to the single field and non-interacting models. In particular, it can fill out regions of $(w_0,w_a)$ parameter space outside the ``thawing quintessence'' regime. This means that interacting models are \emph{less} likely to provide a good explanation for the DESI inference on DE evolution than a single field model. This is further evident in the data space of the distance measures themselves, where adding more fields causes the prior measure to move away from $\Lambda$CDM and from the preferred $(w_0,w_a)$ models. Nevertheless, we studied interesting cases with exotic evolution in detail. The presence of interactions allows for behaviours whereby the fields are oscillating, but the equation of state does not average to zero, and rather has secular evolution in the average that can even become more negative over time. Turns in field space and the onset of strong second third or fourth field evolution complicate this behaviour further. We show in the Appendix how one can explain this behaviour in a rotating eigenbasis model yielded possible by the interaction terms.

We only considered the prior probability on distance measures for multifield axion models. Computing the posterior probability on the parameters is beyond the scope of this work. However, the possibility remains that exotic field dyanmics, such as multiple turns and oscillations, could be relevant in explaining future anomalous data. The danger in such a case would be overfitting of noisy data, and so care should be taken with the statistical method \citep[see e.g.][for relevant methods]{Ong:2026tta}.

While the interesting dynamics we have found seem to make interacting models less desirable from the point of view of explaining DESI, such behaviours should be generic in the axiverse \citep[related phenomena are explained by the ``friendship'' concept studied for dark matter axions by][]{Cyncynates:2021xzw}. These exotic behaviours will be important to understand further in multi-axion models not just for DE, but also for inflation, reheating, and dark matter relic density, in cases where the dynamics cannot be reduced to single field approximations.

\section*{Acknowledgments}

SK was supported by Mahidol University Scholarship for Undergraduate Student Mobility Program, Fiscal Year 2025. DJEM is supported by ST/T004037/1 and ST/X000753/1 from the Science and Technologies Facilities Council UK. We acknowledge useful discussions with Jakob Moritz.

\appendix

\section{A: Rotating Eigenbasis Model}

To understand how the potential generates the interactions of the axion fields, we consider small field fluctuations and expand $V$:
\begin{align}
    V(\phi_1,\phi_2)&\approx\Lambda_1^4\left(\frac{1}{2!}\left(\frac{Q_{11}\phi_1}{f_1} + \frac{Q_{12}\phi_2}{f_2}\right)^2-\frac{1}{4!}\left(\frac{Q_{11}\phi_1}{f_1} + \frac{Q_{12}\phi_2}{f_2}\right)^4+\mathcal{O}^6\right)\nonumber\\&+\Lambda_2^4\left(\frac{1}{2!}\left(\frac{Q_{21}\phi_1}{f_1} + \frac{Q_{22}\phi_2}{f_2}\right)^2-\frac{1}{4!}\left(\frac{Q_{21}\phi_1}{f_1} + \frac{Q_{22}\phi_2}{f_2}\right)^4+\mathcal{O}^6\right).
\end{align}
This reduces to
\begin{align}\label{approc-pot}
    V(\phi_1,\phi_2)\approx\phi_iM^2_{ij}\phi_j-\lambda_{klnm}\phi_k\phi_l\phi_n\phi_m +\mathcal{O}(\phi^6)
\end{align}
where $M_{ij}$ is the mass matrix and $\lambda_{klnm}$ is the quartic coupling constant tensor~\citep{Mehta:2021pwf}.

The equations of motion are:
\begin{align}
    \ddot\phi_1+3H\dot\phi_1+\frac{\Lambda_1^4}{f_1}\sin\left(\frac{\phi_1}{f_1}+\frac{\phi_2}{f_{12}}\right)&=0\, ,\label{int-eom1} \\
    \ddot\phi_2+3H\dot\phi_2+\frac{\Lambda_1^4}{f_{12}}\sin\left(\frac{\phi_1}{f_1}+\frac{\phi_2}{f_{12}}\right)+\frac{\Lambda_2^4}{f_2}\sin\left(\frac{\phi_2}{f_2}\right)&=0\, . \label{int-eom2}
\end{align}
Taylor expanding:
\begin{align}
    \ddot\phi_1+3H\dot\phi_1+\frac{\Lambda_1^4}{f_1}\left\{\left(\frac{\phi_1}{f_1}+\frac{\phi_2}{f_{12}}\right)+\frac{1}{3!}\left(\frac{\phi_1}{f_1}+\frac{\phi_2}{f_{12}}\right)^3+\mathcal{O}^5\right\}&\approx0\, , \\
    \ddot\phi_2+3H\dot\phi_2+\frac{\Lambda_1^4}{f_{12}}\left\{\left(\frac{\phi_1}{f_1}+\frac{\phi_2}{f_{12}}\right)+\frac{1}{3!}\left(\frac{\phi_1}{f_1}+\frac{\phi_2}{f_{12}}\right)^3+\mathcal{O}^5\right\}+\frac{\Lambda_2^4}{f_2}\left\{\left(\frac{\phi_2}{f_2}\right)+\frac{1}{3!}\left(\frac{\phi_2}{f_2}\right)^3+\mathcal{O}^5\right\}&\approx0\, .
\end{align}
At leading order:
\begin{align}
    \ddot{\vec{\phi}}+3H\dot{\vec{\phi}}+M^2\vec{\phi}=0\label{eom-matrix}
\end{align}
where $M^2$ is the mass matrix given by
\begin{align}\label{mass-matrix}
    M^2=\begin{pmatrix}
        \frac{\Lambda_1^4}{f_1^2} & \frac{\Lambda_1^4}{f_1f_{12}}\\
        \frac{\Lambda_1^4}{f_1f_{12}} & \frac{\Lambda_1^4}{f_{12}^2}+\frac{\Lambda_2^4}{f_2^2}
    \end{pmatrix}
\end{align}
Next, we diagonalize the mass matrix by rotating the fields $\vec{\phi}\rightarrow R\vec\phi=\vec\chi$, leading to
\begin{align}
    V(\vec\phi)&\approx\vec\phi ^TM^2\vec\phi-\mathcal{O}^4\, ,\\
    &\approx\vec\chi^TRM^2R^T\chi-\mathcal{O}^4\, , \\
    &=\vec\chi^TM_d^2\vec\chi\, ,
\end{align}
with equations of motion
\begin{align}
    \ddot{\vec{\chi}}+3H\dot{\vec{\chi}}+M_d^2\vec{\chi}=0\, .
\end{align}
By using the WKB approximation with $M_d\gg H$, we obtain
\begin{align}
    \chi_i\approx a^{-3/2}(t)A_i\cos(m_it+\delta_i)
\end{align}
where $i=\pm$ and $m_i$ is the eigenvalue of $M_d^2$ which are given by:
\begin{align}
    m_{\pm}^{2}
=\frac12\Big(m_1^{2}(1+r^{2})+m_2^{2}\Big)
\pm
\frac12\sqrt{\Big(m_1^{2}-m_1^{2}r^{2}-m_2^{2}\Big)^{2}+4\,m_1^{4}r^{2}},\label{mpm}
\end{align}
where $r=f_1/f_{12}$. Here, it can be seen that $m_\pm>m_a$. Since mixing can increase the effective eigenmasses relative to the original masses, the system can enter the oscillatory regime $H\lesssim m_\pm$ earlier than in the single-field or non-interacting case.
For the fields in the original basis, this gives:
\begin{align}\label{sol1}
    \phi_1(t)&=a^{-3/2}A_+\cos\theta\cos(m_+t+\delta_+)+a^{-3/2}A_-\sin\theta\cos(m_-t+\delta_-)
\end{align}
\begin{align}\label{sol2}
    \phi_2(t)&=-a^{-3/2}A_+\sin\theta\cos(m_+t+\delta_+)+a^{-3/2}A_-\cos\theta\cos(m_-t+\delta_-)
\end{align}

Next, we consider the effect of interactions as acting to give a non-constant eigenbasis. Thus:
\begin{align}
    \ddot{\vec{\chi}}+3H\dot{\vec{\chi}}-2\tilde{A}\dot{\vec{\chi}}+(M_d^2-\dot{\tilde{A}}+\tilde{A}^2-3H\tilde{A})\vec{\chi}=0
\end{align}
where
\begin{align}
    \tilde{A}=R\dot{R}^T=\begin{pmatrix}
        0&\dot\theta\\-\dot\theta&0
    \end{pmatrix}
\end{align}
and we redefine the fields as $\vec{u}=a^{3/2}\vec{\chi}$. The equations of motion in terms of the $\vec{u}$ field are
\begin{align}
    \ddot{\vec{u}}-2\tilde{A}\dot{\vec{u}}+\left[M_d-\frac{3}{2}\dot{H}-\frac{9}{4}H^2-\dot{\tilde{A}}+\tilde
    {A}^2\right]\vec{u}=0
\end{align}
that split to two equation of motions as
\begin{align}
    \ddot{u}_++2\dot\theta\dot{u}_-+\omega_+^2+\ddot\theta u_-&=0\\
    \ddot{u}_--2\dot\theta\dot{u}_++\omega_-^2-\ddot\theta u_+&=0
\end{align}
where $\omega^2_\pm=m^2_\pm-\frac{3}{2}\dot{H}-\frac{9}{4}H^2-\dot\theta^2$. 

For the oscillating case in dark energy domination $(m_\pm\gg H,\dot\theta,\dot\omega/\omega)$, we have the condition of slow varying oscillation such that $\epsilon\equiv\frac{H}{\omega}\ll1$ and $\ddot H\approx 0$. We approximate $u_i=A_i e^{-i\int\omega_i dt}+\text{c.c.}$, where $A_i(t)=C_i(t)/\sqrt{2\omega_i}$. With this condition, we obtain the expression of $C_i(t)$ as
\begin{align}
    \dot C_+(t)&= e^{i\int\Delta\omega dt}\left[-\dot\theta\sqrt{\frac{\omega_-}{\omega_+}}-i\frac{\ddot\theta}{2\sqrt{\omega_-\omega_+}}\right]C_-(t)\\
    \dot C_-(t)&=-e^{-i\int\Delta\omega dt}\left[-\dot\theta\sqrt{\frac{\omega_+}{\omega_-}}+i\frac{\ddot\theta}{2\sqrt{\omega_-\omega_+}}\right]C_+(t)
\end{align}
In case of $\omega^2_i\sim m^2_i$, the expression of $\chi_i$ is
\begin{align}
    \chi_\pm\approx a^{-3/2}\cos(m_\pm t+\delta_\pm)\int dt e^{\pm i\int\Delta\omega dt}\left[\mp\dot\theta\sqrt{\frac{\omega_{\mp}}{\omega_\pm}}-i\frac{\ddot\theta}{2\sqrt{\omega_-\omega_+}}\right]C_\mp(t)
\end{align}
and for $\phi_1$ and $\phi_2$ this gives
\begin{align}
\begin{split}
\phi_1(t) \;\approx\; a^{-3/2}\Bigg[ &
\cos\theta \, \cos\!\big(m_+ t + \delta_+\big) 
   \int dt e^{ i\int\Delta\omega dt}\left[-\dot\theta\sqrt{\frac{\omega_{-}}{\omega_+}}-i\frac{\ddot\theta}{2\sqrt{\omega_-\omega_+}}\right]C_-(t) \\
&+ \sin\theta \, \cos\!\big(m_- t + \delta_-\big) 
   \int dt e^{- i\int\Delta\omega dt}\left[\dot\theta\sqrt{\frac{\omega_{+}}{\omega_-}}-i\frac{\ddot\theta}{2\sqrt{\omega_-\omega_+}}\right]C_+(t)
\Bigg] .
\end{split}
\end{align}
\begin{align}
\begin{split}
\phi_2(t) \;\approx\; a^{-3/2}\Bigg[ &
-\sin\theta \, \cos\!\big(m_+ t + \delta_+\big) 
    \int dt e^{ i\int\Delta\omega dt}\left[-\dot\theta\sqrt{\frac{\omega_{-}}{\omega_+}}-i\frac{\ddot\theta}{2\sqrt{\omega_-\omega_+}}\right]C_-(t)  \\
&+ \cos\theta \, \cos\!\big(m_- t + \delta_-\big) 
   \int dt e^{- i\int\Delta\omega dt}\left[\dot\theta\sqrt{\frac{\omega_{+}}{\omega_-}}-i\frac{\ddot\theta}{2\sqrt{\omega_-\omega_+}}\right]C_+(t)
\Bigg] .
\end{split}
\end{align}

These expressions for $\phi_1$ and $\phi_2$ can explain the trajectory in the ($\phi_1,\phi_2$) subspace e.g. Fig.\ref{evo1}b, in which we can see qualitatively the rotation of the basis when it oscillates in the late-time universe ($\log_{10} a\approx 0$). Recall the equation of state is:
\begin{align}
    w_\phi=\frac{P}{\rho}=\frac{\frac{1}{2}\dot\phi_1^2+\frac{1}{2}\dot\phi_2^2-V(\phi_1,\phi_2)}{\frac{1}{2}\dot\phi_1^2+\frac{1}{2}\dot\phi_2^2+V(\phi_1,\phi_2)}
\end{align}
or in the mass eigenbasis,
\begin{align}
    w_\phi&=\left(\frac{\frac{1}{2}\dot\phi_1^2+\frac{1}{2}\dot\phi_2^2-V(\phi_1,\phi_2)}{\frac{1}{2}\dot\phi_1^2+\frac{1}{2}\dot\phi_2^2+V(\phi_1,\phi_2)}\right)_{\phi_i\rightarrow R^T_{ij}\chi_j}\\
    &=\frac{\frac{1}{2}\dot{\vec{\chi}}^T\dot{\vec{\chi}}+\frac{1}{2}\dot\theta^2\vec\chi^T\vec\chi+\dot\theta\dot{\vec{\chi}}^T J\vec\chi-\left(\frac{1}{2}m_+^2\chi_+^2+\frac{1}{2}m_-^2\chi_-^2\right)}{\frac{1}{2}\dot{\vec{\chi}}^T\dot{\vec{\chi}}+\frac{1}{2}\dot\theta^2\vec\chi^T\vec\chi+\dot\theta\dot{\vec{\chi}}^T J\vec\chi+\left(\frac{1}{2}m_+^2\chi_+^2+\frac{1}{2}m_-^2\chi_-^2\right)}
\end{align}
where $J$ is
\begin{align}
    J=\begin{pmatrix}
        0&1\\-1&0\, .
    \end{pmatrix}
\end{align}
The average values of $\omega_\phi$ is thus
\begin{align}
    \langle w_\phi\rangle=\frac{\langle P\rangle}{\langle\rho\rangle}
\end{align}

For each term, we approximate the $C_\pm/m_\pm, H/m_\pm$ as very small and take $\dot\theta^2$ to vary slowly such that $\dot\theta^2(t+\delta t)\approx\dot\theta^2(t)+\delta\dot\theta^2$ where $\delta\dot\theta^2/\theta^2\ll1$. Thus, $\dot\theta^2\approx const.$ for a short period of time. Then, we use the condition of slow varying oscillation, and we obtain
\begin{align}
    \langle\dot\chi^2_\pm\rangle&\approx \frac{1}{2}a^{-3}m_\pm^2\frac{\abs{C_\pm}^2}{\omega_\pm}\\
    \langle\chi^2_\pm\rangle&\approx\frac{1}{2}a^{-3}\frac{\abs{C_\pm}^2}{\omega_\pm}\\
    \langle\dot{\vec{\chi}}^TJ\vec\chi\rangle&=0
\end{align}
Then, we obtain $\langle\omega_\phi\rangle$ as
\begin{align}\label{avg-w}
    \langle w_\phi\rangle&\approx\frac{\dot\theta^2\sum\abs{C_\pm}^2/\omega_\pm}{2\sum\abs{C_\pm}^2m_\pm^2/\omega_\pm+\dot\theta^2\sum\abs{C_\pm}^2/\omega_\pm}\\
    &\approx \frac{\dot\theta^2}{2M_{eff}}
\end{align}
where we define $M_{eff}$ as $M_{eff}\equiv\frac{\sum\abs{C_\pm}^2m_\pm^2}{\sum\abs{C_\pm}^2}$. From Eq. \ref{avg-w}, the average of equation of state depends on time via $(\dot\theta^2)$ which slowly decreases (see Fig.\ref{evo2}b and \ref{fig:3a2}, lower panels), and the average value of equation of state also decreases with time (see the bottom right panel of each figure). 

\begin{figure}
    \centering
    \includegraphics[width=0.33\linewidth]{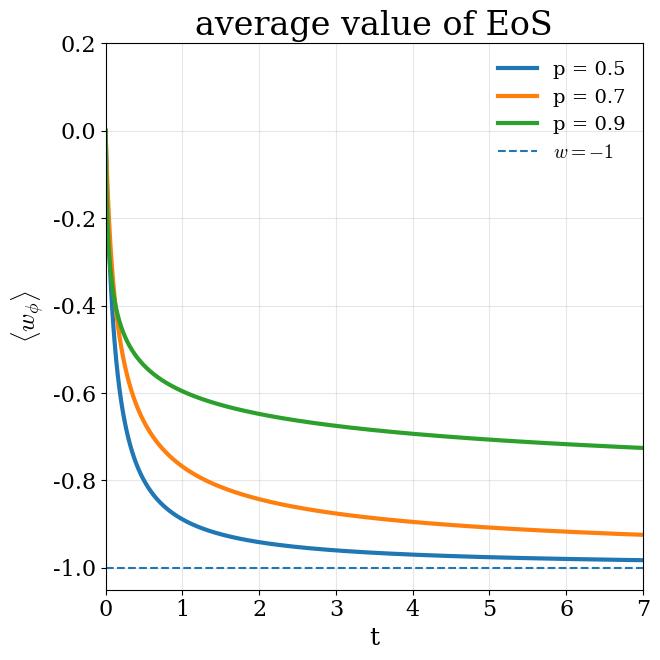}
    \includegraphics[width=0.33\linewidth]{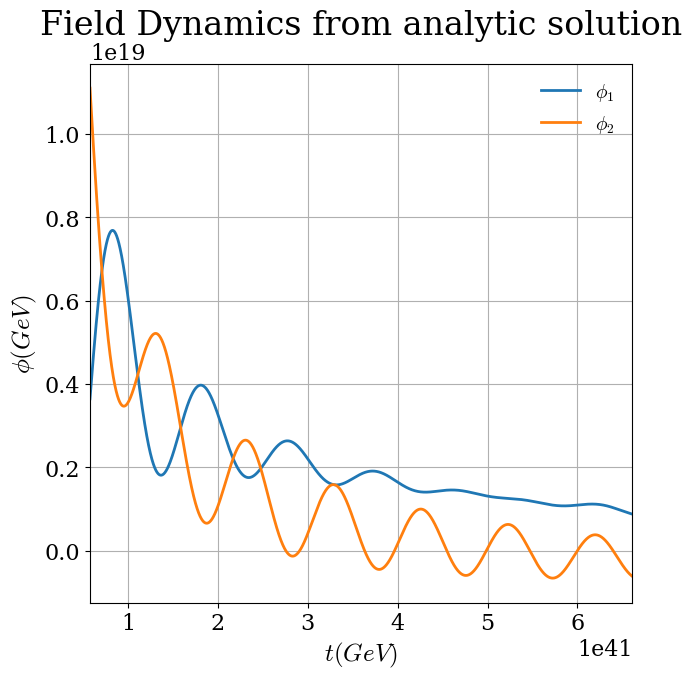}
    \includegraphics[width=0.33\linewidth]{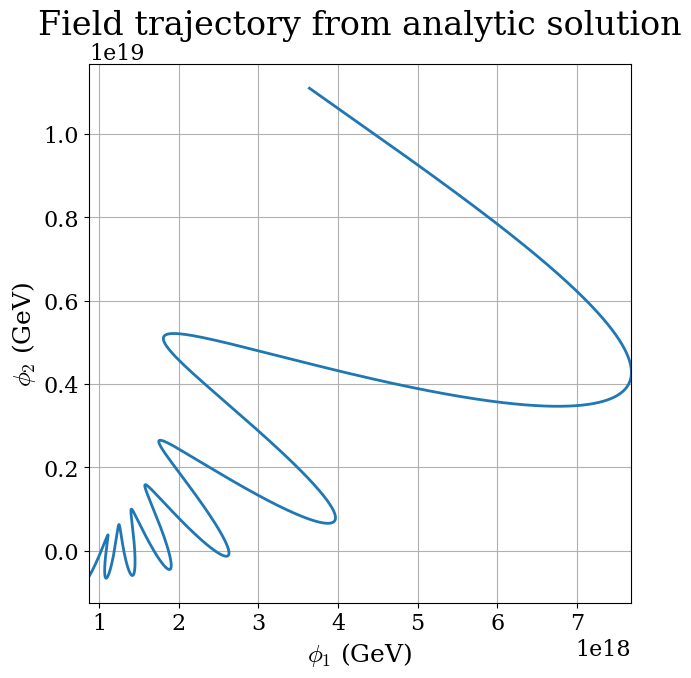}
    \caption{\emph{Left}: $\langle w_\phi\rangle$ at different values of $p$, which paramaterises the slow evolution of the eigenbasis by angle $\theta\propto t^p$. \emph{Middle}: Evolution of $\phi_1$ and $\phi_2$ from \eqref{sol1}, and \eqref{sol2}. \emph{Right:} Trajectory of $\phi_1$ and $\phi_2$.}
    \label{fig:w_p_theta_model}
\end{figure}

In the same way, we can find the average of equation of state by using the expression of $\phi_1$ and $\phi_2$ in Eq. \ref{sol1}, and Eq. \ref{sol2}, respectively. By taking the rotation angle to depend on time, we will get the same form of $\langle\omega_\phi\rangle$ as
\begin{align}
    \langle w_\phi\rangle&\approx\frac{\dot\theta^2\sum A_\pm^2}{2\sum A_\pm^2m_\pm^2+\dot\theta^2\sum A_\pm^2}\\
    &\approx\frac{\dot\theta^2}{2M_{eff}^2},
\end{align}
where $M_{eff}\equiv\frac{\sum A_\pm^2m_\pm^2}{\sum A_\pm^2}$ which is the same form as in a previous method. Next, taking the ansatz of $\theta(t)$ as $ct^p$, we can write $\langle\omega_\phi\rangle$ as
\begin{align}
    \langle w_\phi\rangle\approx\frac{c^2p^2t^{2(p-1)}}{2M_{eff}}
\end{align}
which will decrease in time if $p<1$ and we shift the function to 

\begin{align}
    \langle w_\phi\rangle\approx-1+\frac{c^2p^2t^{2(p-1)}}{2M_{eff}}
\end{align}
which we plot in Fig.~\ref{fig:w_p_theta_model} (left). In that plot, we have shifted $t\to t+\left(\frac{2M_{\rm eff}}{c^{2}p^{2}}\right)^{\frac{1}{2(p-1)}}$ to be consistent with $\langle w_\phi\rangle=0$ at the initial time. Taking $\theta(t)=ct^p$ with the condition is $0<p<1$, and setting the initial condition is $\dot\phi_i(t_0)=0$,  we can obtain the coefficient $A_\pm$ as $A_\pm=a^{3/2}\phi_i/\cos\delta_\pm$. The evolution of $\phi_1$, and $\phi_2$ are shown in Fig.~\ref{fig:w_p_theta_model} (middle and right). 
\section{B: Analytical Analysis for N-fields multi-axion quintessence}\label{analy-N}
In this section, we will illustrate the analytical analysis for N-field axion quintessence. For the N-field model, we recall the axion potential from Eq. \eqref{N-potential}. \\
\indent The equations of motion are:
\begin{gather}
\ddot{\phi}_1
+ 3H\dot{\phi}_1
+ m_1^2 f_1
\sin\left(
\frac{\phi_1}{f_1}
+ \frac{\phi_2}{f_{12}}
+ \cdots
+ \frac{\phi_N}{f_{1N}}
\right)
= 0 ,
\\
\ddot{\phi}_2
+ 3H\dot{\phi}_2
+ \frac{m_1^2 f_1^2}{f_{12}}
\sin\left(
\frac{\phi_1}{f_1}
+ \frac{\phi_2}{f_{12}}
+ \cdots
+ \frac{\phi_N}{f_{1N}}
\right)
+ m_2^2 f_2
\sin\left(
\frac{\phi_2}{f_2}
+ \frac{\phi_3}{f_{23}}
+ \cdots
+ \frac{\phi_N}{f_{2N}}
\right)
= 0 ,
\\
\vdots\nonumber\\
\ddot{\phi}_N
+ 3H\dot{\phi}_N
+ \sum_{a=1}^{N}
\frac{m_a^2 f_a^2}{f_{aN}}
\sin\left(
\sum_{j=a}^{N}
\frac{\phi_j}{f_{aj}}
\right)
= 0 .
\end{gather}
Taylor expanding, and keeping only leading order, we obtain:
\begin{gather}
\ddot{\phi}_1
+ 3H\dot{\phi}_1
+ m_1^2 f_1
\left(
\frac{\phi_1}{f_1}
+ \frac{\phi_2}{f_{12}}
+ \cdots
+ \frac{\phi_N}{f_{1N}}
\right)+\mathcal{O}^3
= 0 ,
\\
\ddot{\phi}_2
+ 3H\dot{\phi}_2
+ \frac{m_1^2 f_1^2}{f_{12}}
\left(
\frac{\phi_1}{f_1}
+ \frac{\phi_2}{f_{12}}
+ \cdots
+ \frac{\phi_N}{f_{1N}}
\right)
+ m_2^2 f_2
\left(
\frac{\phi_2}{f_2}
+ \frac{\phi_3}{f_{23}}
+ \cdots
+ \frac{\phi_N}{f_{2N}}
\right)+\mathcal{O}^3
= 0 ,
\\
\vdots\nonumber\\
\ddot{\phi}_N
+ 3H\dot{\phi}_N
+ \sum_{a=1}^{N}
\frac{m_a^2 f_a^2}{f_{aN}}
\left(
\sum_{j=a}^{N}
\frac{\phi_j}{f_{aj}}
\right)+\mathcal{O}^3
= 0 .
\end{gather}
Writing this system of equation of motion in the matrix form as Eq. \eqref{eom-matrix}, the mass matrix for N-field system is given by:
\begin{equation}
M^2 =
\begin{pmatrix}
m_1^2
&
m_1^2\dfrac{f_1}{f_{12}}
&
\cdots
&
m_1^2\dfrac{f_1}{f_{1N}}
\\[0.35cm]
m_1^2\dfrac{f_1}{f_{12}}
&
m_1^2\dfrac{f_1^2}{f_{12}^2}+m_2^2
&
\cdots
&
m_1^2\dfrac{f_1^2}{f_{12}f_{1N}}
+
m_2^2\dfrac{f_2}{f_{2N}}
\\
\vdots
&
\vdots
&
\ddots
&
\vdots
\\
m_1^2\dfrac{f_1}{f_{1N}}
&
m_1^2\dfrac{f_1^2}{f_{1N}f_{12}}
+
m_2^2\dfrac{f_2}{f_{2N}}
&
\cdots
&
\displaystyle
\sum_{a=1}^{N}
\frac{m_a^2 f_a^2}{f_{aN}^2}
\end{pmatrix}.
\end{equation}
Here, we diagonalize this mass matrix using the matrix representation of $SO(N)$. For example, with 3 axions we can rotate to the mass eigenbasis using a matrix of the form:
\begin{align}
    R^{(3)}(\alpha,\beta,\gamma)&\equiv R_{z}(\alpha)R_{y}(\beta)R_{z}(\gamma)\\
    &=\begin{pmatrix}
\cos\alpha\cos\beta\cos\gamma-\sin\alpha\sin\gamma
&
-\cos\alpha\cos\beta\sin\gamma-\sin\alpha\cos\gamma
&
\cos\alpha\sin\beta
\\
\sin\alpha\cos\beta\cos\gamma+\cos\alpha\sin\gamma
&
-\sin\alpha\cos\beta\sin\gamma+\cos\alpha\cos\gamma
&
\sin\alpha\sin\beta
\\
-\sin\beta\cos\gamma
&
\sin\beta\sin\gamma
&
\cos\beta
\end{pmatrix}.
\end{align}
The solution takes the form:
\begin{subequations}\label{eq:three_axion_wkb_solution}
\begin{multline}
\phi_1(t)
=
a^{-3/2}(t)
\Big[
\left(\cos\alpha\cos\beta\cos\gamma-\sin\alpha\sin\gamma\right)
A_+\cos(m_+t+\delta_+)
\\
+
\left(\cos\alpha\cos\beta\sin\gamma+\sin\alpha\cos\gamma\right)
A_-\cos(m_-t+\delta_-)
+
\cos\alpha\sin\beta
A_0\cos(m_0t+\delta_0)
\Big],
\end{multline}
\begin{multline}
\phi_2(t)
=
a^{-3/2}(t)
\Big[
\left(-\sin\alpha\cos\beta\cos\gamma-\cos\alpha\sin\gamma\right)
A_+\cos(m_+t+\delta_+)
\\
+
\left(-\sin\alpha\cos\beta\sin\gamma+\cos\alpha\cos\gamma\right)
A_-\cos(m_-t+\delta_-)
-
\sin\alpha\sin\beta
A_0\cos(m_0t+\delta_0)
\Big],
\end{multline}
\begin{align}
\phi_3(t)
=
a^{-3/2}(t)
\Big[
-\sin\beta\cos\gamma
A_+\cos(m_+t+\delta_+)
-
\sin\beta\sin\gamma
A_-\cos(m_-t+\delta_-)
+
\cos\beta
A_0\cos(m_0t+\delta_0)
\Big],
\end{align}
\end{subequations}
where $m_+, m_-$, and $m_0$ are the eigenvalues of the $3\times 3$ mass matrix. Next, taking time-dependent mixing angle $\alpha,\beta,\gamma$ (i.e. power law time-dependent mixing angle, such as $\alpha\propto t^{p}$), can lead to rotating eigenbasis behaviour such as seen in the top panel of Fig. \ref{fig:3a2}.\\
 \indent For larger \(N\), we use the \(N\times N\) matrix representation of
the rotation group \(SO(N)\) \citep{osti_4683974}. After diagonalizing the
mass matrix, the original field basis and the mass eigenbasis are related by
an orthogonal rotation matrix \(R\in SO(N)\). Therefore, the solution for an
\(N\)-field system can be written as:
\begin{equation}
\begin{pmatrix}
\phi_1(t)\\
\phi_2(t)\\
\phi_3(t)\\
\vdots\\
\phi_N(t)
\end{pmatrix}
=
R
\begin{pmatrix}
\chi_1(t)\\
\chi_2(t)\\
\chi_3(t)\\
\vdots\\
\chi_N(t)
\end{pmatrix},
\end{equation}
where
\begin{equation}
R^T M^2 R
=
M_d^2
=
\operatorname{diag}
\left(
m_1^2,m_2^2,m_3^2,\ldots,m_N^2
\right),
\end{equation}
$m_i$ is the eigenvalue, and $\chi_i(t)$ is the solution in the mass eigenbasis.
\section{C: Benchmarks Comparison}\label{sec:C}
In this section, we will examine the distance measurement comparison from the examples from the Section~\ref{sec:dynamics}.\\
\begin{figure}
    \centering
    \includegraphics[width=\linewidth]{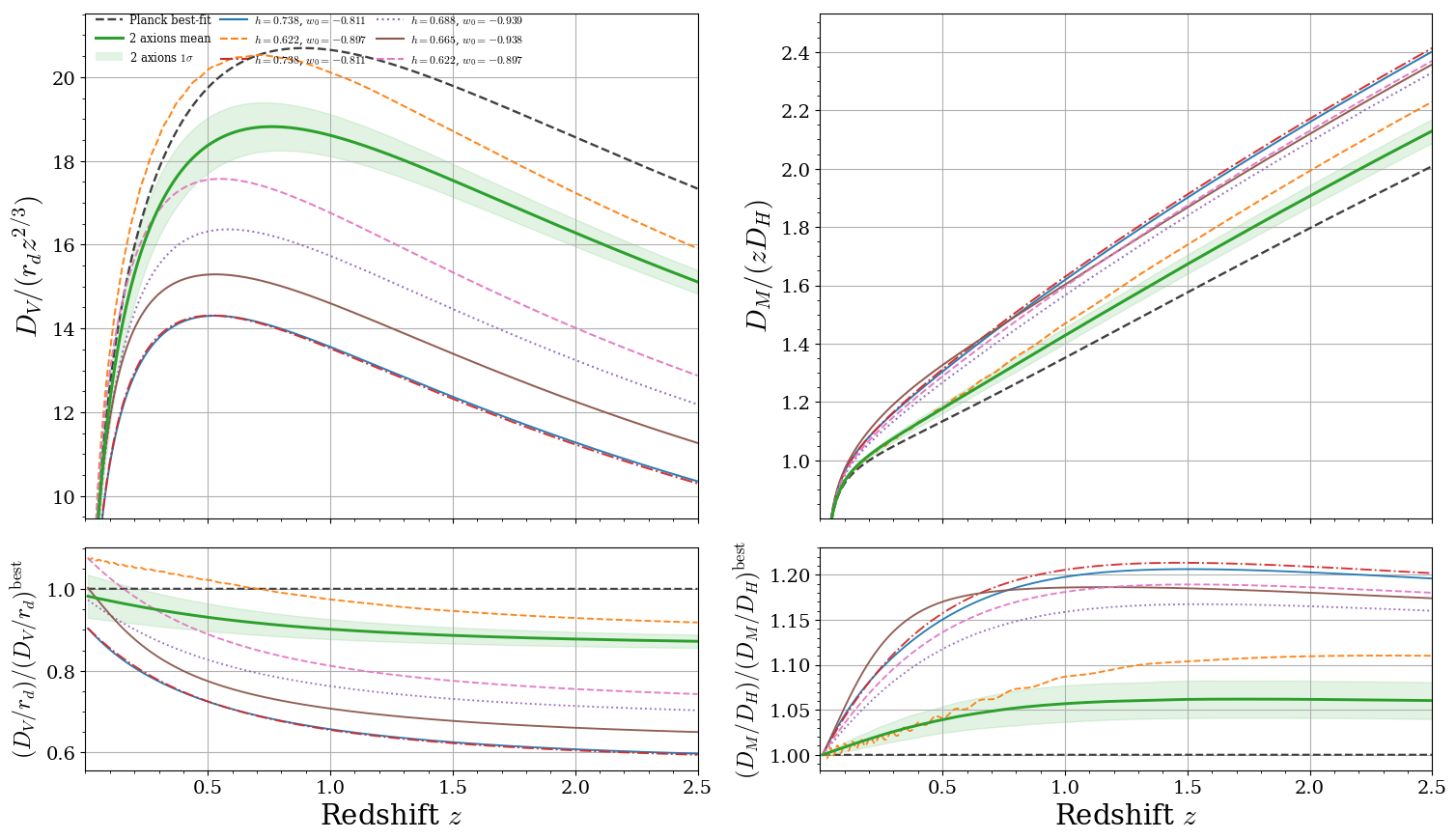}
    \caption{Distance measures from selected exotic two axion models.}
    \label{fig:2a-compare}
\end{figure}
\begin{figure}
    \centering
    \includegraphics[width=\linewidth]{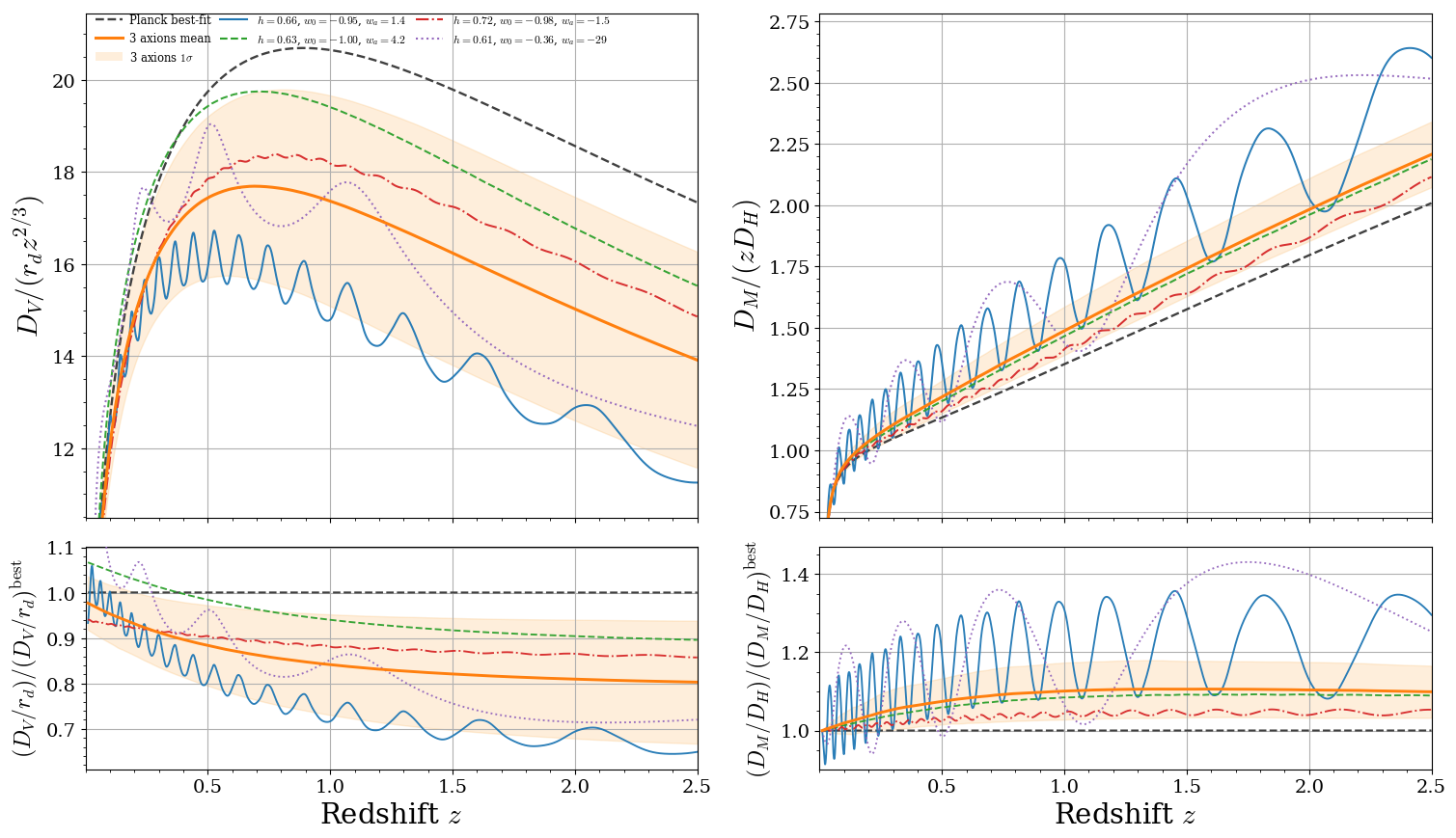}
    \caption{Distance measures from selected exotic three axion models.}
    \label{fig:3a-compare}
\end{figure}
\begin{figure}
    \centering
    \includegraphics[width=\linewidth]{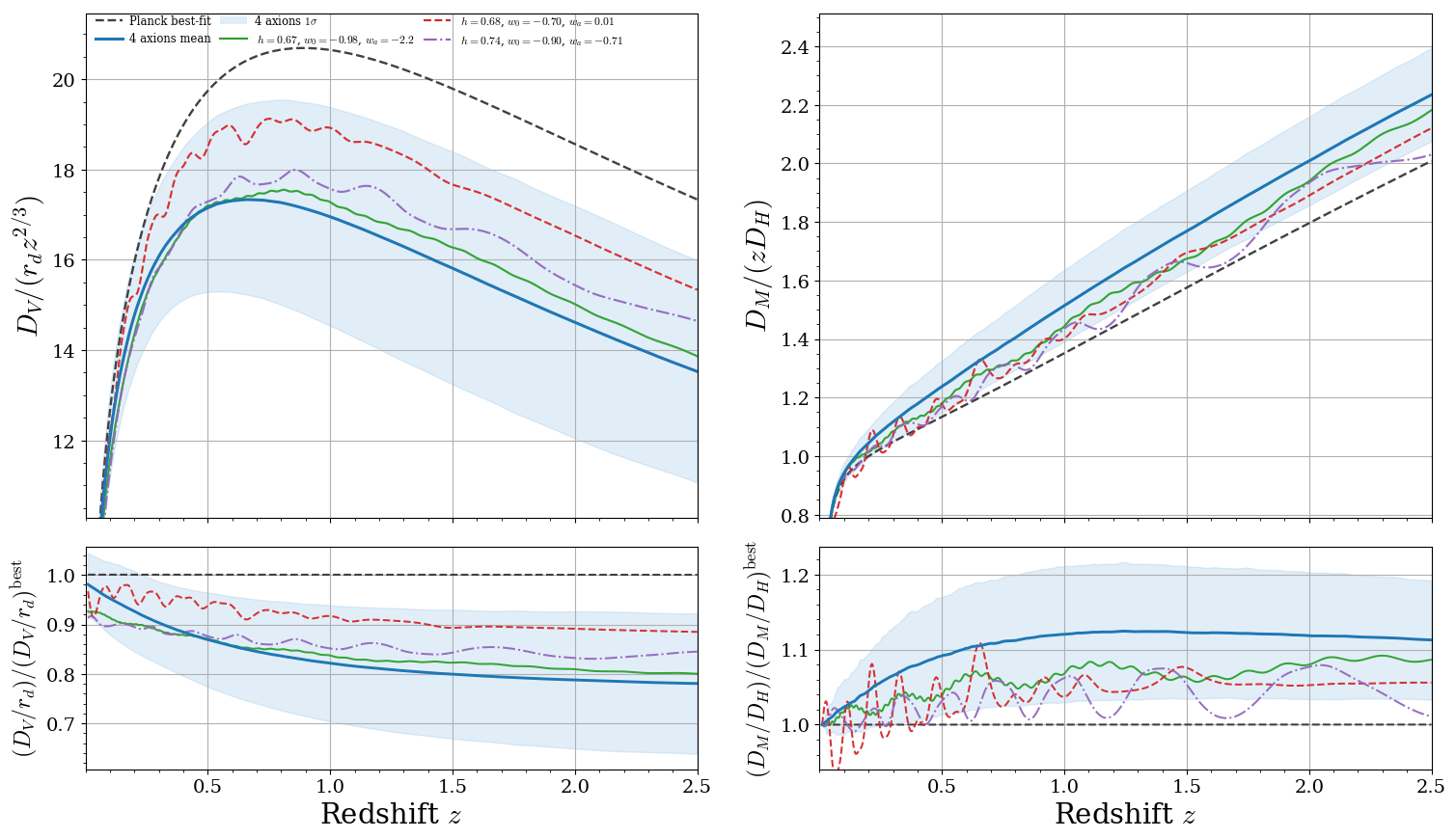}
    \caption{Distance measures from selected exotic four axion models.}
    \label{fig:4a-compare}
\end{figure}
 Figure~\ref{fig:2a-compare} shows the benchmark comparison of cosmic distances in the 2-axion quintessence model. Different initial conditions lead to visible shifts in the distance measures at high redshift, especially in the top-left and bottom-left panels. In the top-left panel, three benchmark cases yield substantially smaller values of $D_V/(r_d z^{2/3})$ compared with the Planck best-fit curve and the 2-axion mean. Similarly, the bottom-left panel shows that all benchmark models satisfy $(D_V/r_d)/(D_V/r_d)^{\rm best}<1$, indicating a systematic suppression of the BAO distance relative to the Planck best-fit prediction. On the other hand, the right-hand panels show larger values for all benchmark models, implying an enhancement of the ratio $D_M/D_H$ compared with the Planck best-fit result.\\
\indent Figure~\ref{fig:3a-compare} illustrates the benchmark comparison of cosmic distance in the 3-axion quintessence model. In this figure, there are 2 benchmarks which experience non-stable cosmic distances. These behaviour are caused by strong oscillation of the axion energy density (e.g. see the energy density panel of the top panel of Figure~\ref{fig:3a1}, and the bottom panel of Figure~\ref{fig:3a2}). In the left-hand panels, the cosmic distances are generally suppressed relative to the Planck best-fit prediction, with the oscillatory benchmarks fluctuating around or below the 3-axion mean. In contrast, the right-hand panels show oscillatory behaviour above both the 3-axion mean and the Planck best-fit curve. Therefore, the 3-axion quintessence model can produce richer and less smooth distance evolution than the 2-axion case, mainly due to the stronger oscillatory behaviour of the axion fields.\\
\indent Figure~\ref{fig:4a-compare} presents the benchmark comparison of cosmic distances in the 4-axion quintessence model. From the selected examples, we still observe oscillatory behaviour in the cosmic distances, similar to that found in the 3-axion quintessence model. This feature is caused by the oscillating energy density of the axion sector, especially for the red dashed and purple dash-dotted benchmarks. These benchmarks also show a suppression in the ratio $(D_V/r_d)/(D_V/r_d)^{\rm best}$, while the ratio $(D_M/D_H)/(D_M/D_H)^{\rm best}$ is enhanced, as in the 3-axion quintessence model.

\bibliographystyle{apsrev4-1}

\bibliography{mybib}

\end{document}